\documentclass[acmsmall,nonacm,screen]{acmart}
\usepackage{xspace}
\usepackage{xcolor,colortbl}
\usepackage{listings}
\usepackage{multirow}
\usepackage{syntax}
\PassOptionsToPackage{dvipsnames,table}{xcolor}

\AtBeginDocument{%
  }

\acmISBN{978-1-4503-XXXX-X/18/06}

\author{Elizabeth Gilbert}
\email{evgilber@andrew.cmu.edu}
\orcid{0009-0005-7242-6179}
\affiliation{
  \institution{Carnegie Mellon University}
  \city{Pittsburgh}
  \state{Pennsylvania}
  \country{USA}
}
\author{Matthew Schneider}
\orcid{0009-0009-4266-7210}
\affiliation{
  \institution{Carnegie Mellon University}
  \city{Pittsburgh}
  \state{Pennsylvania}
  \country{USA}
}
\author{Zixi An}
\orcid{0009-0004-2557-6230}
\affiliation{
  \institution{Carnegie Mellon University}
  \city{Pittsburgh}
  \state{Pennsylvania}
  \country{USA}
}
\author{Suhas Thalanki}
\orcid{0009-0003-5631-2767}
\affiliation{
  \institution{Carnegie Mellon University}
  \city{Pittsburgh}
  \state{Pennsylvania}
  \country{USA}
}
\author{Wavid Bowman}
\orcid{0009-0000-9228-591X}
\affiliation{
  \institution{Florida Institute for National Security}
  \city{Gainesville}
  \state{Florida}
  \country{USA}
}
\author{Alexander Bai}
\orcid{0009-0009-7458-7864}
\affiliation{
  \institution{Max Planck Institute for Software Systems}
  \city{Saabrucken}
  \state{Saarland}
  \country{Germany}
}
\author{Ben L. Titzer}
\orcid{0000-0002-9690-2089}
\affiliation{
  \institution{Carnegie Mellon University}
  \city{Pittsburgh}
  \state{Pennsylvania}
  \country{USA}
}
\author{Heather Miller}
\orcid{0000-0002-2059-5406}
\affiliation{
  \institution{Carnegie Mellon University}
  \city{Pittsburgh}
  \state{Pennsylvania}
  \country{USA}
}

\renewcommand{\shortauthors}{Gilbert et al.}

\usepackage[utf8]{inputenc}

\begin{document}

\newcommand{\todo}[1]{\textbf{\textcolor{red}{TODO: #1 }}}
\newcommand{\evg}[1]{\textbf{\textcolor{teal}{EVG: #1 }}}
\newcommand{\orca}{Seal\xspace}
\newcommand{\ourlang}{Whamm\xspace}
\newcommand{\oureng}{\texttt{whamm}\xspace}
\newcommand{\ourcomp}{Whamm compiler\xspace}
\newcommand{\Ourparam}{Whamm\mbox{-}\texttt{param}\xspace}
\newcommand{\Ourparams}{\texttt{Vec\textlangle Whamm\mbox{-}param\textrangle}\xspace}
\newcommand{\ourparam}{\texttt{whamm}\mbox{-}\texttt{param}\xspace}
\newcommand{\ourarg}{\texttt{whamm}\mbox{-}\texttt{arg}\xspace}
\newcommand{\ourargs}{\texttt{whamm}\mbox{-}\texttt{args}\xspace}
\newcommand{\ourmonitor}{\texttt{whamm}\mbox{-}\texttt{monitor}\xspace}
\newcommand{\ourprobe}{\texttt{whamm}\mbox{-}\texttt{probe}\xspace}
\newcommand{\ourfunc}{\texttt{whamm}\mbox{-}\texttt{function}\xspace}
\newcommand{\ourprobes}{\texttt{whamm}\mbox{-}\texttt{probes}\xspace}
\newcommand{\strue}{\textcolor{teal}{\texttt{true}}\xspace}
\newcommand{\sfalse}{\textcolor{teal}{\texttt{false}}\xspace}
\newcommand{\X}{$\times$}

\definecolor{comment}{HTML}{78909c}
\definecolor{keyword}{HTML}{ed7d3c}
\definecolor{special}{HTML}{4371c6}

\title{Debugging WebAssembly? Put some \ourlang on it!}

\begin{abstract}
Debugging and monitoring programs are integral to engineering and deploying software.
Dynamic analyses monitor applications through source code or IR injection, machine code or bytecode rewriting, and virtual machine or direct hardware support.
While these techniques are viable within their respective domains, common tooling across techniques is rare, leading to fragmentation of skills, duplicated efforts, and inconsistent feature support.
We address this problem in the WebAssembly ecosystem with \ourlang, a declarative instrumentation DSL for WebAssembly that abstracts above the instrumentation strategy, leveraging bytecode rewriting and engine support as available.
\ourlang solves three problems: 1) tooling fragmentation, 2) prohibitive instrumentation overhead of general-purpose frameworks, and 3) tedium of tailoring low-level high-performance mechanisms.
\ourlang provides fully-programmable instrumentation with declarative match rules, static and dynamic predication, automatic state reporting, and user library support, while achieving high performance through compiler and engine optimizations.
At the back end, \ourlang provides instrumentation to a Wasm engine as Wasm code, reusing existing engine optimizations and unlocking new ones, most notably \emph{intrinsification}, to minimize overhead.
In particular, explicitly requesting program state in match rules, rather than reflection, enables the engine to efficiently bundle arguments and even \emph{inline} compiled probe logic.
\ourlang streamlines the tooling effort, as its bytecode-rewriting target can run instrumented programs everywhere, lowering fragmentation and advancing the state of the art for engine support.
We evaluate \ourlang with case studies of non-trivial monitors and show it is expressive, powerful, and efficient.
\end{abstract}

\maketitle

\section{Introduction}

\begin{figure*}
    \includegraphics[width=1.00\linewidth]{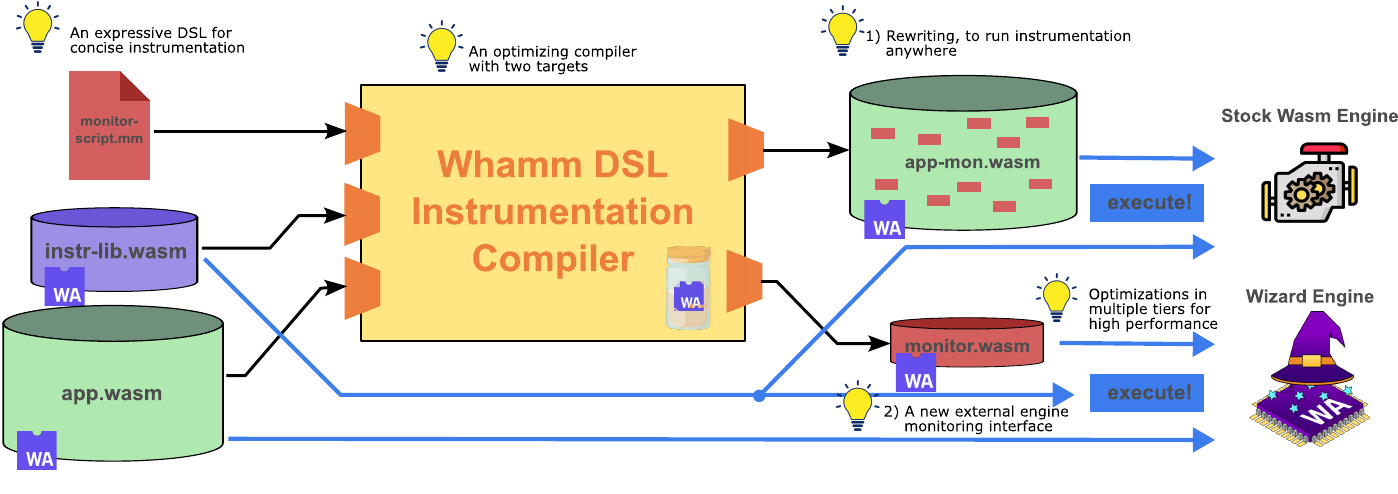}
    \caption{Contribution map of this paper. We have designed and implemented a new expressive DSL (codenamed \ourlang) for concise instrumentation of Wasm applications with support for extensible libraries. It features an optimizing compiler that targets either rewriting (which injects bytecode into the application and can run anywhere) or a new engine monitoring interface (codenamed \oureng) with high performance.}
    \label{fig:contributions}
\end{figure*}

For decades, software tooling has supported developers in debugging and optimizing their programs.
While static analysis can provide many definitive insights about programs, only in running and observing programs through a \emph{dynamic analysis} do we truly understand them.
Tools outside of a program, such as the operating system kernel, libraries, and profilers can observe events like page faults, cache misses, branch prediction misses, I/O events, and signals.
These tools can report event statistics and memory usage without additional instrumentation.
To gather more specific information, programs can be instrumented at the source level (either manually or via the compiler), at the bytecode/IR level, at the emulator/VM level, or with dynamic binary rewriting.

We use the term \emph{monitor} to refer to a self-contained analysis that observes an application and reports on its behavior, such as execution patterns, internal states, and resource consumption.
Monitors may \emph{instrument} a program by many strategies, modifying the program before or during execution, logging desired events, then generating a report after the program has terminated.
Ideally monitors are \emph{non-intrusive}; that is, they should not alter the semantic behavior of the program.
Of course, any instrumentation may \emph{perturb} performance by increasing execution time or other resource usage.
A dynamic analysis tool should minimize overhead.

\subsection{Instrumentation}

\textbf{Source-level instrumentation.} It is possible to directly write monitor code within or alongside source code, but this has a number of drawbacks.
Instrumentation may be unable to capture low-level operations such as I/O accesses, system calls, and signals.
It creates cross-cutting concerns as source and monitor code are mixed and increases the risk that the monitor is \emph{intrusive}.
Often, the program incurs some performance hits as frequent checks are required.

\textbf{Rewriting.}
With rewriting, new instructions are inserted directly into the application at points of interest.
These new instructions collect, summarize, and report various types of information gathered at runtime.
Rewriting is usually performed programmatically, instead of manually, either at the source, bytecode, or machine-code level.

\emph{Static rewriting} involves injecting monitor code into the source code before compilation, or into bytecode or machine code before execution. 
This rewriting is decoupled from any particular execution platform, increasing portability and expressibility of the monitors.
However, the inserted code may inadvertantly alter application behavior, even innocently.
On native targets, the presence of additional code can alter program state, since a program may read its own code as data.
With bytecode and machine code rewriting, the binary needs to be reorganized to fit the monitoring code, which may not always be possible, and source-level mappings need to be adjusted to handle the new code.
Static rewriting tools include Oron~\cite{Oron} (at the source level) and EEL~\cite{EEL} (at the machine code level).

In \emph{dynamic rewriting}, monitoring code is inserted at runtime, which can avoid some of the limitations of static rewriting. 
For example, when an application dynamically generates new code, static rewriting might not instrument it, but dynamic rewriting can.
However, dynamic rewriting requires more sophisticated tools, e.g. dynamic recompilation, which also means that additional instrumentation cost is paid at runtime.
Dynamic rewriting tools include Jalangi~\cite{Jalangi} (at the source level) and DTrace~\cite{DTrace} and Pin~\cite{Pin} (at the machine code level).

Rewriting can be tedious and is not portable across ISAs.
Composing multiple monitors is difficult, and most rewriting frameworks do not automatically handle the necessary parallel composition, as sequential composition would lead to one monitor instrumenting another.

\textbf{Emulation.}
Emulators and VMs such as QEMU~\cite{QEMU} and Valgrind~\cite{Valgrind} track instrumentation points separately from the application code, requiring no rewriting or re-organization for better developer ergonomics.
Instrumentation is invoked outside of the emulated state space, which prevents altering behavior.
However, even with purpose-built JITs for emulation, this strategy incurs additional overheads due to unstructured machine code and is generally considered prohibitively expensive for heavy instrumentation.

\subsection{WebAssembly}

WebAssembly (Wasm)~\cite{WasmPldi}, is a portable, low-level bytecode format that serves as an efficient target for many programming languages, such as C, C++, and Rust.
While initially designed to run programs on the Web, due to its sandboxed execution environment, it has gained adoption in other contexts, such as cloud computing~\cite{CloudflareWasm}, edge computing~\cite{FastlyEdge}, and embedded systems~\cite{WasmIndMach}.
Wasm engines offer interpreter and compiler tiers, which together achieve quick startup times and high peak performance.
Yet despite its spread and the sophistication of its engines, no standard instrumentation, profiling, or debugging interface for Wasm exists.

\textbf{State of Wasm instrumentation.}
Today, static rewriting of bytecode is the norm for Wasm instrumentation, but tooling is low-level and clunky.
Several static instrumentation tools are available: BREWasm~\cite{brewasm}, AspectWasm~\cite{AspectWasm}, Walrus~\cite{Walrus}, and Wasabi~\cite{Wasabi}.
These tools are less mature than rewriting tools for platforms such as the JVM.
Compared to native targets, however, static rewriting on the WebAssembly platform is less intrusive because the execution model features a Harvard architecture, a virtualized execution stack, and multiple memories.
It is also portable, as Wasm abstracts above the host ISA.
Wasm engines can also minimize the impact of certain overheads; JIT tiers can speed up checks on probes using adaptive optimizations.
However, instrumenting bytecode is still tedious because injected code has to account for application semantics and deal with Wasm's stack machine architecture.

Alternatively, with direct support in the virtual machine, e.g. Wizard~\cite{WizardInstr}, bytecodes can be instrumented without rewriting, which simplifies composability of instrumentation and makes it completely non-intrusive.
Both the interpreter and compiler tiers of a Wasm engine can benefit additionally from optimizations specific to instrumentation.
One such class of optimizations is called \emph{intrinsification} in which the engine handles certain cases specially, e.g. by executing special logic in the interpreter and emitting specialized machine code for counter probes in the JIT.

\subsection{Contributions}

We illustrate the components of the \ourlang system in Figure~\ref{fig:contributions} and explain our contributions here.
This paper introduces the first expressive, retargetable DSL, codenamed \ourlang, designed specifically for Wasm instrumentation, and two complete high-performance realizations of the system based on bytecode rewriting and novel engine support.
With bytecode rewriting, the \ourlang compiler can inject instrumentation directly into application code, resulting in an instrumented program that can run on any engine.
Alternatively, when targeting an \emph{engine interface}, the \ourlang compiler generates a \emph{monitor module} that can be loaded into a Wasm engine and reused across applications.
To support high-performance instrumentation directly in the Wasm engine, we extend the Wizard Research Engine~\cite{WizardEngine} with the ability to load monitors expressed as Wasm modules.
Whereas previously, monitors could only be expressed as extensions that require rebuilding the engine, monitor modules are more flexible and extensible.
We show a number of \ourcomp optimizations, including partial evaluation of predicates, as well as engine optimizations such as custom trampolines and inlining instrumentation probes directly into the application code.
In particular, we contribute:

\begin{itemize}

\item
\textbf{A fully-featured declarative instrumentation DSL for Wasm, codenamed \ourlang.}
We present \ourlang, a domain-specific language for instrumenting Wasm code that is expressive and performant.
Our work is open source~\footnote{URL omitted for anonymity.} and will be provided as an artifact.
We demonstrate its language features with a suite of complex monitors.

\item
\textbf{The first DSL that abstracts above the instrumentation strategy.}
\ourlang can either rewrite bytecode, allowing monitors to run on any Wasm engine, or interface with the engine directly, which can be more composable and efficient.

\item
\textbf{A suite of compiler optimizations that reduces overhead.}
We show how the design of \ourlang admits optimizations in both the bytecode-rewriting and engine strategies to reduce the overhead of instrumentation both statically and dynamically.

\item
\textbf{An engine monitoring interface, codenamed \oureng, based on monitor modules.}
We show how Wasm engines can support extensible dynamic instrumentation by loading ordinary Wasm modules that conform to a monitor interface.

\item
\textbf{A suite of engine optimizations that increases performance for \oureng.}
We show how Wasm engines can optimize both interpreter and compiler tiers, achieving performance competitive with that only previously available via special-case optimizations.

\item
\textbf{An empirical evaluation on standard benchmarks and workloads.}
We demonstrate the effectiveness of \ourlang's language features and optimizations through a detailed evaluation of several complex monitors, comparing bytecode rewriting and direct engine support with similar dynamic analysis tools like Pin and Wasabi.
Our detailed analysis isolates individual overheads in various tiering configurations to study the effectiveness of each optimization.

\end{itemize}

\section{Instrumentation with \ourlang}

\ourlang is a domain-specific language for WebAssembly instrumentation that solves the long-standing problem of either suffering poor performance, poor portability, or poor composability.
It enables developer tooling while reducing hassle since it abstracts \emph{above the instrumentation technique} to best leverage the capabilities available at runtime.
\ourlang can instrument applications using the \emph{bytecode-rewriting} technique or by \emph{interfacing with an engine} that directly supports instrumentation. 
Together, our language supports a broad domain of applications without requiring a developer to reimplement their monitors to satisfy any particular instrumentation technique or engine.

\subsection{The \ourlang Language}\label{subsec:the-language}

\begin{figure*}
    \includegraphics[width=.80\linewidth]{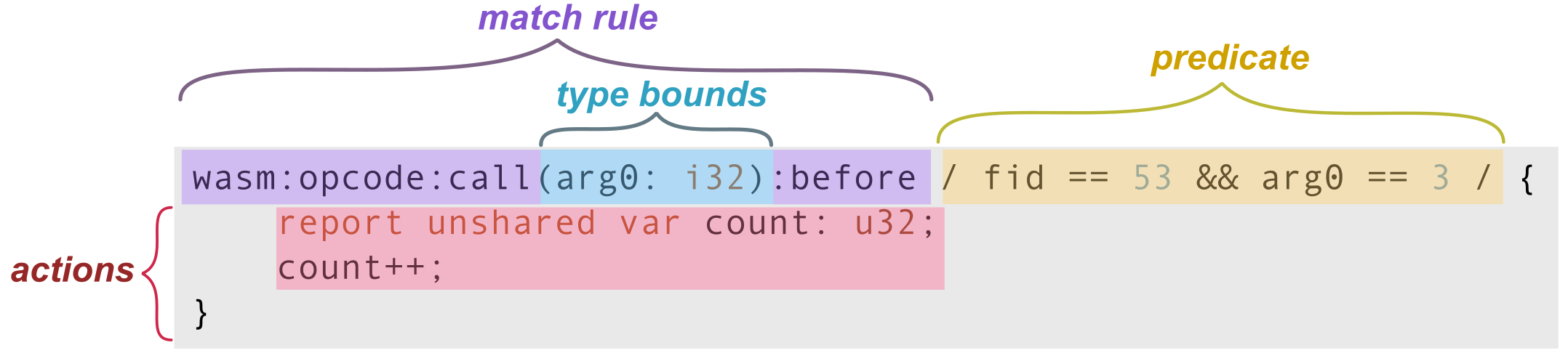}
    \caption{Directives in \ourlang consist of a triple of a \emph{match rule} that includes bounds on typed arguments, an optional \emph{predicate}, and consequent \emph{actions}.}
    \label{fig:mm-example-with-labels}
\end{figure*}

The syntax of \ourlang is inspired by Dtrace~\cite{DTrace}.
Users can express their instrumentation as \emph{directives} that reference program \emph{events} and provide predicated \emph{actions}, as shown in Figure~\ref{fig:mm-example-with-labels}.
This form can be read as, ``When this \emph{event} occurs during program execution, do these \emph{actions} if this \emph{predicate} is \strue.''
A directive provides a high-level and intuitive syntax that specifies events at various granularities in the target program.
For example, a developer can instrument a function exit either by targeting the \texttt{wasm:function:exit} event provided by the DSL or by explicitly handling each bytecode that may return. 

\subsubsection{Match Rules}\label{subsubsec:match-rules}
The match rule part of the directive constitutes a syntactic pattern matching one or more program events.
It consists of four parts: \texttt{provider}, \texttt{package}, \texttt{event}, and \texttt{mode}, described in Table~\ref{tab:match-rules}.
Each part of the match rule decreases in specificity until reaching the \texttt{mode} of the probe. 
As an example, consider the following match rule:
\begin{lstlisting}[escapechar=!]
!\hspace{20pt}\textcolor{keyword}{wasm}!:!\textcolor{keyword}{opcode}!:!\textcolor{keyword}{br_if}!:!\textcolor{keyword}{before}!
\end{lstlisting}
This rule can be read as, ``Insert this probe \emph{before} each of the \emph{\texttt{br\_if} Wasm opcodes} in the program.''
Match rules support wildcards to enable matching on a group of events in the same manner.
For example, the match rule
\begin{lstlisting}[escapechar=!]
!\hspace{20pt}\textcolor{keyword}{wasm}!:!\textcolor{keyword}{opcode}!:!\textcolor{special}{*}\textcolor{keyword}{load}\textcolor{special}{*}!|!\textcolor{special}{*}\textcolor{keyword}{store}\textcolor{special}{*}!:!\textcolor{keyword}{before}!
\end{lstlisting}
matches all variants of the load and store Wasm opcodes.

It is possible to modify program behavior through leveraging the \emph{alt} probe \texttt{mode}.
This \texttt{mode} replaces the original logic at some target bytecode location.
Currently, this feature is only fully supported by the \ourlang rewriting target.

\begin{table}
    \centering
    \begin{tabular}{p{0.15\linewidth} | p{0.8\linewidth}}
        Part & Description \\
        \hline\hline
        \texttt{provider} & The \texttt{provider} that supports the capability used in the probe. \\
        \hline
        \texttt{package} & The \texttt{package} within the provider that supports the capability used in this probe. \\
        \hline
        \texttt{event} & The \texttt{event} describing the location(s) for probe insertion. \\
        \hline
        \texttt{mode} & The \texttt{mode} used to emit probe actions at target location(s), e.g. \texttt{before} and \texttt{after}. \\
        \hline
    \end{tabular}
    \caption{A match rule in \ourlang consists of four parts.}
    \label{tab:match-rules}
\end{table}

\textbf{Type Bounds}.
Some Wasm opcodes, such as \texttt{call}, \texttt{local.set}, and \texttt{select} are \emph{polymorphic}, meaning their inputs and outputs can be different types depending on the function or module context.
Variables in a probe body that refer to these arguments have types that depend on the match site.
One way to address this would be to introduce polymorphism or dynamic typing.
Instead, to avoid dynamic typing, we require programs to use \emph{type bounds}, where the user expresses the expected types of each argument in scope to the probe body when there is more than one static possibility.
A type mismatch is then considered a \emph{no match} for the rule\footnote{For type bounds that could never be satisfied, such as matching on an \texttt{i64} input to an \texttt{i32.add} instruction, the compiler emits an error.}.
Thus, correct matching requires knowing the types of all arguments at all program points.

\subsubsection{Predication}\label{subsubsec:predication}
A programmer can instrument events under specific conditions with \emph{predicates}, expressions that must evaluate to \strue for a probe to be executed.
An empty predicate is always \strue.
A probe can be predicated on either static information, such as the match location within a function (\texttt{pc}), or dynamic information, such as the instruction's arguments, or a mix of the two in a compound logical expression.
The \ourlang compiler implements predication differently depending on the target.
For bytecode rewriting, the \ourlang compiler performs constant propagation and folding for statically-known variables such as the \texttt{pc} and function id (\texttt{fid}).
If any residual predicate remains, the compiler injects instrumentation on the remaining dynamic part.
For the engine target, it will automatically split the predicate into static and dynamic parts, with the dynamic part compiled into the probe actions, as explained in Section~\ref{subsubsec:predicate-opts}.
In Figure~\ref{fig:mm-example-with-labels}, the matched \texttt{call} opcodes are statically constrained to those located in function 53 (specified by the static predicate \texttt{fid == 53}) with an \texttt{i32} as its first argument (specified by the type bound on \texttt{arg0}).
The body of the probe will be wrapped in code to check the dynamic part of the predicate (\texttt{arg0 == 3}).
When injecting an \emph{alt} probe, dynamic predication guards the consequent actions with the original logic in the \texttt{else} block.

%

\subsubsection{Storage Classes}\label{subsubsec:storage-classes}
A strikingly useful feature of \ourlang is a unique set of storage classes for variables that simplify managing the lifetime and reporting of collected data.
For example, variables can be \emph{global} or \emph{local}, \emph{shared} across multiple matches, or shared \emph{per execution frame}.

\textbf{Bound functions and globals.}\label{bound-globals}
Data specific to the event, such as the operands to an instruction, the location of the match, or the name of a target function in a call are directly available in the probe body as \emph{bound variables}.
Each part of the match-rule hierarchy binds appropriate functions and globals.
For example, the \texttt{wasm:opcode:call} event provides a \texttt{target\_fn\_name} variable that can be used in the probe directive to refer to a function's name in the name section.
While some variables are statically-known for a match location, other variables are dynamic.
In \ourlang, all arguments to an instruction (i.e. values on the Wasm operand stack) are available as named, typed variables \texttt{arg0}, \texttt{arg1}, etc.
To aid users, we have implemented a nifty CLI tool to query the provided functions and variables in scope for a probe directive.

\textbf{Derived variables.}
Having useful, relevant variables in scope automatically, along with match rule wildcards, can greatly simplify writing bytecode-level monitors.
Sometimes such variables aren't inherently part of an event, but are easily computable from the other variables in the event.
We call these \emph{derived variables}.
Consider a monitor which matches on all load and store Wasm opcodes.
It can use the derived \texttt{addr} variable in the probe body to refer to the dynamic memory address.
The \texttt{addr} variable is available for memory-access events and hides the fact that for loads, the target address is stored in the first argument, \texttt{arg0}, while for stores, this value is stored in the second argument, \texttt{arg1}.
Other examples of derived variables include: \texttt{taken}, which for any branching opcode represents whether the branch will be taken~\footnote{For \texttt{br\_if}, \texttt{taken = arg0 != 0}, but the new \texttt{br\_on\_cast} and \texttt{br\_on\_null} opcodes from Wasm-GC require emulating a cast in the probe, which is not easy. Derived variables reduce duplicated work.}; \texttt{effective_addr}, which for loads and stores represents the \texttt{index + offset}; and \texttt{trap}, which indicates whether the instruction \emph{will trap}.

\textbf{Unshared variables}.
Many instrumentation tasks requiring retaining information that is specific to a match site and should be maintained across invocations, such as a counter.
In \ourlang, a programmer may mark a variable as \texttt{unshared}, which means an instance of this variable is allocated for \emph{each individual match site} of the match rule.
The lexical scope of this variable is limited to the probe's body, but the variable will retain its value across multiple invocations of the body.
This can be used to collect data at a specific program point over time.
For example, the following script collects the number of times that \emph{a specific} \texttt{call} opcode was reached during application execution:
\begin{lstlisting}[escapechar=!]
!\hspace{20pt}\textcolor{keyword}{wasm}:\textcolor{keyword}{opcode}:\textcolor{keyword}{call}:\textcolor{keyword}{before}! {
	!\hspace{20pt}\textcolor{keyword}{unshared var} count: \textcolor{special}{u32}!;
	!\hspace{20pt}!count++;
!\hspace{20pt}!}
\end{lstlisting}

\textbf{Shared variables}.
Some instrumentation tasks are more easily expressed if variables are instantiated once \emph{across} match sites.
In a \ourlang probe body, a programmer may mark a variable as \texttt{shared}, which means only one instance of this variable is allocated and is made available at \emph{every match site}.
Like unshared variables, the lexical scope of this variable is limited to the probe's body, and the variable retains its value across multiple invocations of the body, but it is not replicated per site.
A small change to the previous script collects the number of times that \emph{any} \texttt{call} opcode was executed dynamically.
\begin{lstlisting}[escapechar=!]
!\hspace{20pt}\textcolor{keyword}{wasm}:\textcolor{keyword}{opcode}:\textcolor{keyword}{call}:\textcolor{keyword}{before}! {
	!\hspace{20pt}\textcolor{keyword}{shared var} count: \textcolor{special}{u32}!;
	!\hspace{20pt}!count++;
!\hspace{20pt}!}
\end{lstlisting}

\textbf{Report variables}.
Obviously, the purpose of instrumentation is to observe and \emph{report}.
To make this quick and easy, \ourlang allows programmers to annotate any variable with the \texttt{report} keyword to request its value be reported on exit.
The compiler will automatically generate appropriate printing code, flushing a textual representation of the report variables, along with their match locations.
Monitor state is printed to the console or appended to a specified file\footnote{In future versions, this behavior will be made configurable through the user providing the behavior as a Wasm module.} as a comma-separated value listing (CSV) that includes match locations.
CSVs allow for easy analysis by other tools.
To flush the results of the previous \texttt{count} example, simply add the \texttt{report} keyword to the variable:
\begin{lstlisting}[escapechar=!]
!\hspace{20pt}\textcolor{keyword}{wasm}:\textcolor{keyword}{opcode}:\textcolor{keyword}{call}:\textcolor{keyword}{before}! {
	!\hspace{20pt}\textcolor{keyword}{report shared var} count: \textcolor{special}{u32}!;
	!\hspace{20pt}!count++;
!\hspace{20pt}!}
\end{lstlisting}

\textbf{Frame variables}. 
Some variables are bound to the lifetime of a function activation.
For example, profilers often record the time when a function is entered and the time when it is exited.
Handling this properly with recursive functions and exceptions is surprisingly tricky and would otherwise require emulating the callstack.
Instead, \ourlang makes this easy with the \texttt{frame} keyword, which marks variables that should live within a single activation of the target function.
With recursion, multiple copies of the variable co-exist simultaneously, and probe invocations get the appropriate copy for their activation.
For the bytecode-rewriting target, the obvious (and efficient) implementation strategy is to simply introduce a new local variable to the function.
To support frame variables in the engine target, we modified the engine with the ability to append to a function's local variable declarations dynamically.

\subsubsection{User Libraries}
A common problem in DSLs is \emph{feature creep} where more and more general-purpose language features are added, compromising its original focus, becoming a frustratingly underpowered language with limited usefulness.
Rather than risk feature creep, we instead see \ourlang as a way to \emph{weave} instrumentation logic into a program, or to generate a monitor that can allow an engine to do that for us, similar to aspect-oriented programming.
Thus, we can specialize the language itself towards instrumentation expressiveness rather than general-purpose programming.

But how can we scale up to handle big instrumentation tasks?
The answer is \emph{libraries}.
\ourlang allows monitors to use libraries provided as Wasm code which the compiler automatically links in, relegating the heavy lifting of complex data structures and libraries to a user's preferred language.
This extensibility prevents the DSL from becoming unwieldy.

For example, we leveraged this capability in the implementation of the \texttt{map} type in \ourlang.
The \texttt{map} type provides an efficient general-purpose associative array.
Its implementation is provided by a Wasm module implemented as a Rust \texttt{HashMap}.
Further, the cache-simulator monitor in our evaluation, see Section~\ref{subsec:programmability}, demonstrates how to use a Wasm module as a monitor library.
The cache model is implemented in Rust, compiled to Wasm, then linked with \ourlang.
The \ourlang script just focuses on expressive matching and reporting, while the cache model is separate and can be independently swapped.
We envision that library support will allow users to add powerful online statistical processing and sampling strategies by writing the complex parts in a general-purpose language and then linking them with \ourlang scripts.

\section{Implementing \ourlang}\label{sec:implementation}

We implemented a compiler for \ourlang in approximately 33,000 lines of Rust that parses scripts, performs semantic analyses, implements optimizations, and outputs Wasm supporting two distinct targets: bytecode rewriting and engine support.
In the rewriting target, the user supplies scripts and the application module(s) being instrumented together to the \ourlang compiler.
The compiler then applies all match rules, compiles probe bodies to Wasm, and injects instrumentation with rewriting, creating the final instrumented module.
For the engine target, the user need only supply the \ourlang scripts, and the compiler produces a standalone Wasm module that implements a monitor according to the \oureng monitoring interface.
The phases of compilation are what one would expect from a traditional compiler, but with interesting behaviors that stem from: (1) collecting and analyzing bound variables per directive, (2) planning variable-storage allocation, and (3) matching probe events, only for the rewriting target.

\subsection{Bound Variable Collection, \Ourparams}\label{subsec:bound-var-collection}

As explained in Section~\ref{subsubsec:storage-classes}, bound variables represent static and dynamic data at match locations, like the \texttt{pc} or an operand to an instruction.
For each directive, \ourlang gathers used variables into a \Ourparams.
By knowing exactly what a probe directive needs and communicating this to the implementation strategy, the data transfer overhead can be minimized.

\subsection{Planning for Storage Allocation}\label{subsec:plan-storage-allocation}

Variables in \ourlang can have storage classes such as \texttt{frame}, \texttt{report}, \texttt{shared}, and \texttt{unshared} (see Section~\ref{subsubsec:storage-classes}).
Storage allocation for such variables occurs at different times depending on the target.

With rewriting, the compiler is provided the application bytecode so that it can \emph{statically} identify all match locations and thus allocate storage ahead of time.
In probe bodies, expressions referencing such variables can be reduced to accesses of constant memory addresses.
However, when targeting an engine, the \ourcomp will produce a monitor module that is application-agnostic.
Later, when applying a monitor module to an application module at load time, the engine will evaluate match rules against and attach probe callbacks to the appropriate application locations.
This late-binding of probes requires the engine and monitor module to cooperate to allocate storage for appropriate variables.
As we'll see in more detail later, the engine supports match-time callbacks so the monitor module can invoke logic each time a new match is found in an application.
The \ourcomp emits callbacks that allocate appropriate per-match-site variables and provide their addresses to probe callbacks.
The probe body uses this memory address as the dynamic part of the address and uses the static \texttt{offset} value of a Wasm \texttt{load} and \texttt{store} operation to point to the appropriate variable location within that allocated region.

\subsection{The Rewriting Strategy}\label{subsec:rewriting-strategy}

With bytecode rewriting, the compiler \emph{directly injects} monitoring logic into the application code.

\subsubsection{The injection library: \orca.}
A critical component of the rewriting target is code to load, parse, reflect on, and modify Wasm binary modules.
As part of this work, we developed a new, fully-featured open-source Rust library for instrumenting Wasm code, codenamed \orca.
\orca allows loading application bytecode into an intermediate representation (IR) that can be \emph{decorated}.
We use decorators on functions and opcodes to attach the bytecode-to-inject with directives on \emph{where} the logic should be injected while emitting the application bytecode post-instrumentation.

The most basic opcode-level decorators have three behavior options: \emph{before}, \emph{after}, and \emph{alternate}.
At module-encoding time, the before and after decorators inject the new bytecode at the original opcode.
The alternate decorator injects the new bytecode \emph{instead of} the original to support application \emph{manipulation}.
The function-level decorators have two behavior options: \emph{entry} and \emph{exit}.
At module-encoding time, the entry decorator injects the new bytecode at the top of the instrumented function.
The exit decorator injects the new bytecode at \emph{all} locations that \emph{could} return in the function body.
\orca's bytecode-injection mechanisms enable library consumers to more easily inject new bytecode into a Wasm module without having to consider the lower-level impacts of injection.

\subsubsection{Identifying match locations.}
As shown in Figure~\ref{fig:injection-rewriting}, we traverse the application bytecode to identify match locations, represented as \orca IR.
When the match rule succeeds, the compiler inspects the \Ourparams to define appropriate data and partially evaluate the predicate through constant propagation.
If the predicate reduces to an expression that depends on dynamic information, the compiler will emit code to evaluate the residual expression as a guard before the probe body.

\subsubsection{Emitting probes at match locations.}
\begin{figure*}
    \includegraphics[width=.80\linewidth]{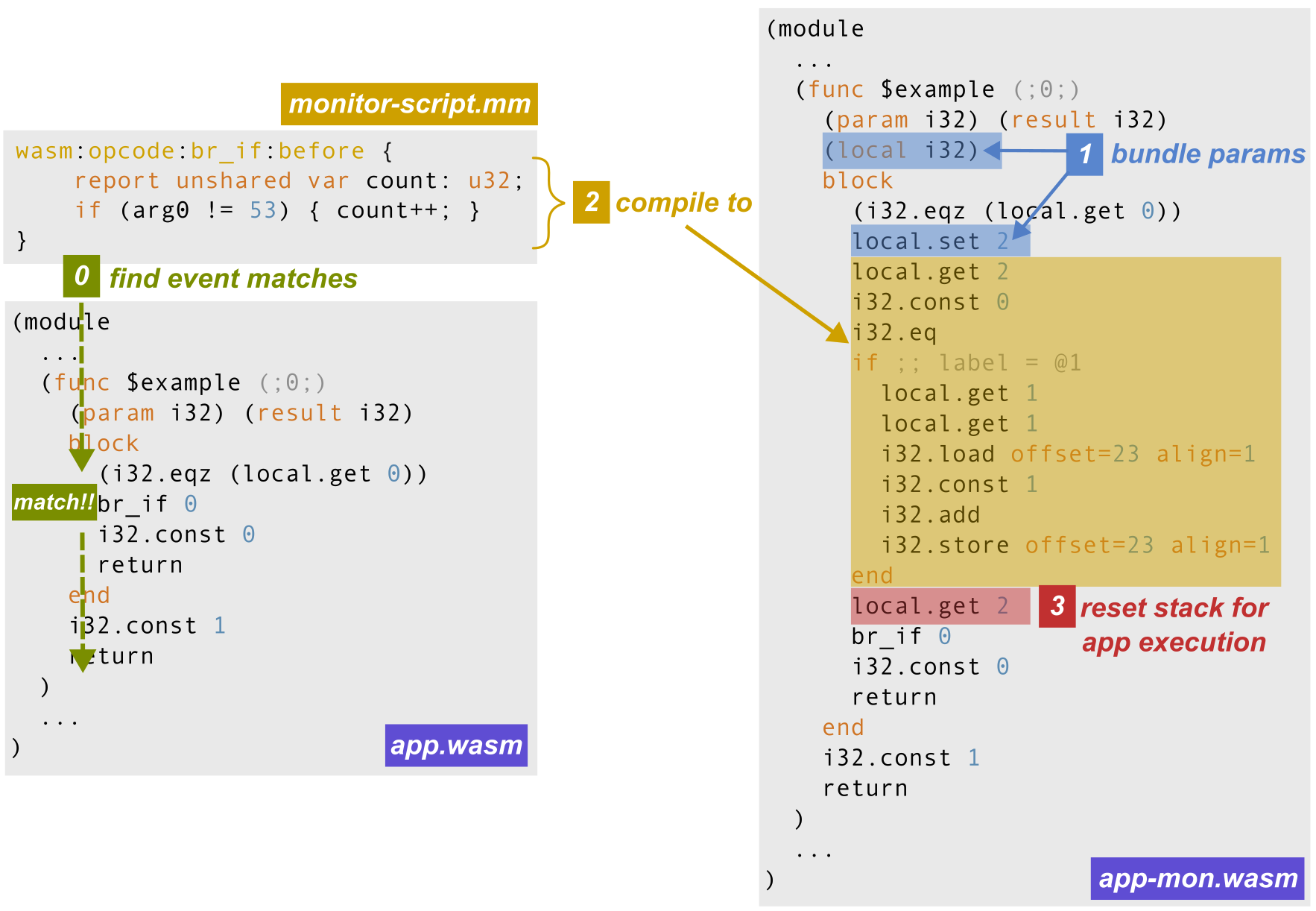}
    \caption{Each step of static instrumentation with \ourlang when targeting the bytecode rewriting.
    In (0) the application module is visited to find event matches, then at a match, (1) the variables referenced by the probe body are saved, (2) the probe body is compiled to WebAssembly and inlined, and (3) the stack is restored to its original state for the original application bytecode.
    }
    \label{fig:injection-rewriting}
\end{figure*}

Because Wasm defines a stack machine, injecting probe logic that uses the input operands to an instruction needs a minor amount of fiddling.
Figure~\ref{fig:injection-rewriting} also shows the three parts of an emitted probe: the parameter prologue, the probe itself, and the stack reset epilogue.
If \Ourparams contains dynamic data requests, these must be bundled as a probe prologue in the emitted bytecode and bound to newly-injected local variables in the function body.
Variables representing operand-stack values are initially saved as locals that can be used in the dynamic predicate and probe body, but cleared later to restore the operand stack.
Our compiler uses local rewrites to avoid reorganizing the operand-stack usage for the entire function.

\subsection{The Engine Interface - \oureng}\label{subsec:engine-strategy}

\begin{figure*}
    \includegraphics[width=.80\linewidth]{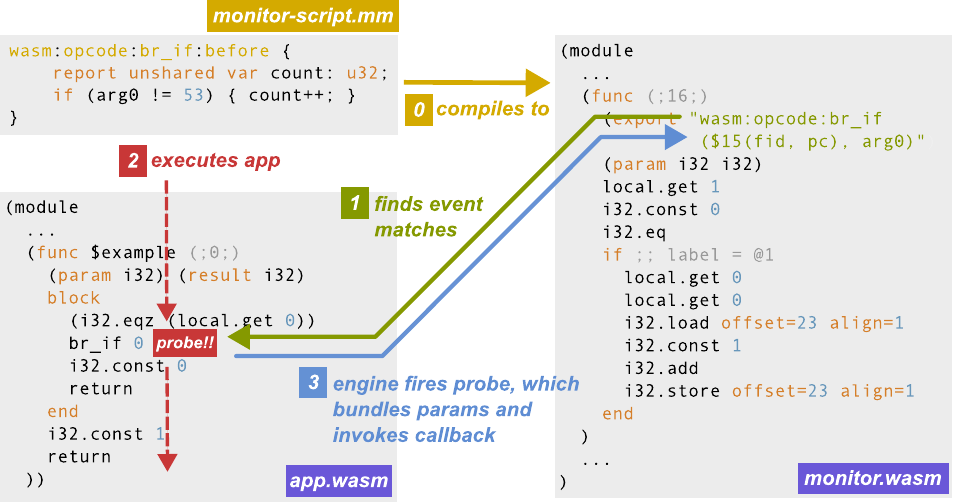}
    \caption{Example of compiling \ourlang to the engine target \oureng.
      The monitor script (0) is compiled to a complete Wasm module that encodes match rules in the names of exported functions.
      At load time, the engine loads the monitor module and (1) applies the match rules to the application-under-observation.
    At runtime (2), when a probed location is reached, the engine will (3) bundle the requested data and invoke the probe callback.
    }
    \label{fig:injection-engine}
\end{figure*}

While \ourlang scripts can run on any engine with the rewriting target, the \ourcomp can also target an engine monitoring interface where the application and monitor are provided separately.
The explicit separation between application and monitor provides additional capabilities and performance improvements over rewriting.
First, preserving the original application code preserves source mappings, making it easier to trace issues identified during instrumentation back to the underlying source.
Second, the engine can enforce separation between monitor state and application state so that a monitor cannot inadvertently alter program behavior.
Third, the engine can compose multiple monitors simultaneously.
Fourth, monitors can be applied more selectively (e.g. per-function) and dynamically added and removed, which unlocks new use cases such as enabling and disabling heavyweight instrumentation to observe rare program behavior.
The engine can completely remove instrumentation when not in use, imposing zero overhead.
Fifth, we can instrument events not observable in bytecode, such as loading/verification/compilation of functions, garbage collection, thread scheduling, or other VM-level operations.

\begin{figure*}
    \includegraphics[width=.75\linewidth]{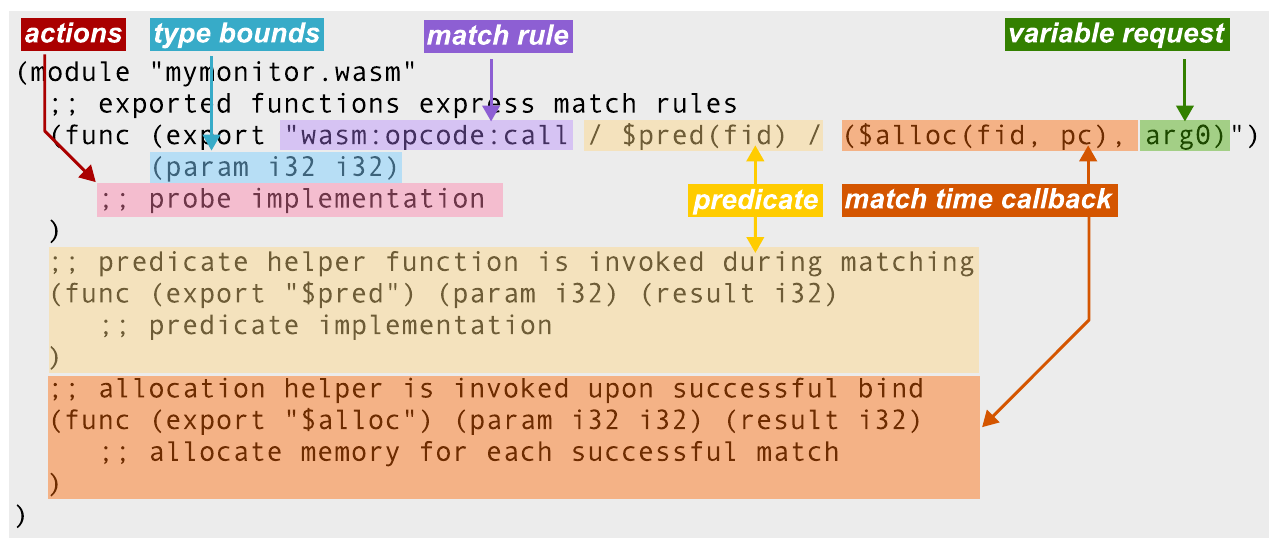}
    \caption{
        Illustrates the \oureng engine interface.
        Wasm monitor modules encode directive \emph{match rule}{s}, \emph{type bounds} and \emph{predicate}{s} into exported functions, expressing where to insert probes declaratively.
    The \emph{match time callback} readies instrumentation state for each match location in the program.
    The \emph{variable request(s)} or {\ourparam}s, instruct the engine on the program state to pass during callback invocation.}
    \label{fig:whamm-engine-interface}
\end{figure*}

\textbf{Monitor modules (\ourmonitor).}
Recently, \cite{WizardInstr} demonstrated the first programmable engine instrumentation for Wasm.
Monitors are written against engine-specific APIs and compiled-in at VM build time, but are not portable across engines.
This has allowed many standard, reusable monitors for Wizard, as well as ad-hoc analyses.
Instead, we discovered that Wasm engines are shockingly good at loading \emph{Wasm modules} and thus the ideal monitor should be provided as a Wasm module.
Thus, we designed an API, where the engine accepts \emph{monitor modules} which provide instrumentation via exported functions.
These regular Wasm modules, referred to as {\ourmonitor}s, conform to a naming convention which is a simplified form of the \ourlang language, called \oureng.
Figure~\ref{fig:whamm-engine-interface} shows an example monitor module that targets the engine interface \oureng.

\textbf{Declarative matching via export names.}
While the rewriting target applies match rules and predicates to a specific application module, there is no application module yet when generating a monitor module.
We require \emph{load-time matching} when applications are loaded.
We considered several design alternatives, such as (1) the Wasm engine provides the \emph{raw bytes} of the application module to the monitor module, which could then re-parse it; (2) the monitor module can \emph{reflect} on the application module through an engine API; or (3) the monitor module communicates matching rules/predicates to the engine declaratively.
We chose the last alternative to reduce duplication between engine and monitor and avoid a costly reparse.
In this scheme, the information necessary for matching---the match rule, static predicate, and runtime data---is encoded in the \emph{export name} of the function as shown in Figure~\ref{fig:whamm-engine-interface}.
The engine parses these export names and performs a simplified form of \ourlang-matching to insert engine-level probes.
For an engine that already supports dynamic instrumentation via probes, the implementation effort required is low (600 lines of code to parse/match/attach/invoke and 250 lines of interpreter and JIT optimizations).

\textbf{Probe and predicate functions.}
The \ourcomp optimizes and translates probe bodies to Wasm functions, called {\ourfunc}s.
Predicates must be evaluated by the engine at match time, so the \ourcomp splits predicates into static and dynamic portions.
Static predicates are wrapped in Wasm functions that can be referred to by their export name in the match rule and dynamic predicates are inlined into the probe bodies, see Section~\ref{sec:pred-splits}.
Both predicates and probes may accept \ourparam{s} that correspond to \ourlang variables, and the engine provides their values (\ourargs) at match time or runtime as necessary.

%

\textbf{Typechecking during matching.}
Some events have polymorphic operands, such as \texttt{select} or \texttt{local.get}.
Type bounds are effectively a type predicate for variables, as described in Section~\ref{subsubsec:match-rules}.
In Wasm bytecode, the types on the operand stack may change from instruction to instruction.
Checking the types of \texttt{argN} variables would therefore seem to be similar to Wasm code validation, a surprisingly tricky endeavor.
Yet, hallelujah, an engine \emph{already has a code validator}!
We extend the code validator with a callback so that the types of \texttt{argN} and \texttt{localN} variables are known at potential probe match sites.
For most rules, a type mismatch is an error, but for type bounds, a mismatch means the probe is simply not inserted.
This is the most complicated part of our engine changes, at approximately 200 lines of code.

\setlength{\grammarindent}{80pt}
\begin{figure}
\begin{grammar}
<export-name> ::= `wasm' `:' `opcode' `:' "OPCODE" <predicate>? <params>?
        \alt `wasm' `:' `exit'

<predicate> ::= `/' <call> `/'

<call> ::= `$' "ID" <params>?

<params> ::= `(' `)'
        \alt `(' <param> ( `,' <param> )* `)'

<param> ::= `argN' | `immN' | `localN' | <call> | `pc' | `fid' | `frame'

\end{grammar}
	\caption{The grammar for \oureng export names in a monitor module.}
	\label{fig:monitor-module-grammar}
\end{figure}

\textbf{Match time callbacks.}
Many monitors store per-match-site data, such as \texttt{unshared} variables in \ourlang.
To implement this functionality and more, the engine interface allows \emph{match-time callbacks} that are invoked by the engine when a match rule and its static predicate are \strue.
A \ourmonitor encodes match-time callbacks into the export name per the grammar shown in Figure~\ref{fig:monitor-module-grammar}.
Statically-known {\ourarg}s, like the function id (\texttt{fid}) and program counter (\texttt{pc}), are allowed as arguments to match-time callbacks.
The \ourmonitor can, for example, allocate a new entry in its internal monitoring data structure and return the entry's address in the \ourmonitor's memory.
The engine passes the result of the static callback---a constant---to the probe function.
Thus one export can define a match rule with a single probe function that accepts custom data for each match site.
The \ourcomp uses this functionality to implement storage for \texttt{unshared} variables.

\textbf{Start function and exit callback.}
In addition to matching on instruction patterns, the engine interface also allows the normal Wasm \texttt{start} function.
Similarly, to report the result of the analysis, the engine will invoke an exported function named "wasm:exit" upon application termination.

\textbf{Summary: an adoptable engine interface.}
The engine interface retains a simplified form of declarative matching expressed in export names but uses Wasm functions for almost everything else: probes, predicates, and reporting logic.
This is to minimize engine burden in hopes of future adoption by more engines.
We chose to prototype the engine interface in the Wizard Research Engine, which already supports a set of bytecode-level dynamic probing mechanisms.
Using these existing mechanisms, we implemented the engine interface in just 600 lines of new code.
With an already well-encapsulated \texttt{Monitor} extension point, extending the Wizard engine to support \ourlang was as simple as writing a new subclass that loads a \ourmonitor, parses its export names, applies match rules, and inserts custom probes that invoke Wasm functions and reporting at exit, all without changing any existing code.
This provided an unoptimized baseline to which we later added interpreter, JIT, and runtime optimizations that we detail in the next section.
In particular, implementing {\ourarg}s naively uses a relatively expensive reflective access to the caller's frame in the probe code.
Nevertheless, given the low implementation burden, we believe that other engines could adopt the same engine interface and thus {\ourmonitor}s \emph{could one day} work on any engine.

\section{Optimizing \ourlang}\label{sec:optimizations}

In this section, we detail optimizations designed to reduce the overhead of instrumentation with \ourlang.
These optimizations fall into two main categories: static optimizations applied by the \ourcomp, and dynamic optimizations implemented in the engine.

\subsection{\ourlang Compiler Optimizations}\label{subsec:whamm-opts}\label{subsubsec:predicate-opts}

Optimizations can be applied by the \ourcomp for both the rewriting and engine targets.

\textbf{Constant propagation.}
Constants appear surprisingly often in \ourlang expressions, since match rules, predicates, and probe bodies often mention bound variables such as \texttt{pc} that become constants for a given match site.
The \ourcomp employs constant propagation to emit better instrumentation code and to optimize predicates.
Deciding more predicates statically means fewer probes will be inserted, and partially evaluating them minimizes the complexity of dynamic predicates.

\textbf{Minimizing operand stack shuffling.}
The \ourcomp instruments Wasm bytecode by rewriting in a mostly linear pass.
When probes use instruction operands (i.e. \texttt{argN}), some stack shuffling is required.
To avoid a global rewrite of an entire function (e.g. reallocating local variables or reshuffling the operand stack), the compiler instead emits the minimal operand-stack save-and-restore code before emitting instrumentation inline.
With \texttt{argN} meaning the $N$-th operand, and \texttt{arg0} referring to the top of stack, the number of operands that must be saved is equal to $M + 1$, $M$ being the largest $N$ used in the directive.

\textbf{Static/dynamic splitting of predicates.}\label{sec:pred-splits}
Match predicates in \ourlang can involve both static and dynamic variables, while in the engine interface, static predicates may be expressed as part of export names.
To map \ourlang predicates onto the engine monitoring interface, the simplest implementation strategy is to classify all predicates as dynamic and emit code within probe bodies to evaluate them.
While correct, this results in instrumenting more match sites and imposing more overhead than necessary on each probe invocation.
A somewhat smarter strategy is to conservatively classify predicates as either fully static or fully dynamic.
This is the strategy the current compiler uses.

However, in the future we can do better by classifying subexpressions as \textcolor{teal}{static}, which are known at match time, and \textcolor{red}{dynamic}, which are only known at runtime, and applying partial evaluation.
For example, consider a static variable \textcolor{teal}{A} and a dynamic variable \textcolor{red}{B} in the predicate expression $p = $ \textcolor{teal}{A} $||$ \textcolor{red}{B}.
By definition, static variables are known before dynamic variables, so we can split the evaluation of an expression into two stages.
In the first stage, static expressions are evaluated, and to derive expressions for the second stage, we substitute into the original expression all the possible outcomes and then apply constant propagation, folding, and strength reduction to obtain a residual dynamic predicate.
To see this, let's assume \textcolor{teal}{A} evaluates to either \strue or \sfalse.
We can then generate two different \emph{residual} dynamic predicates, one where \textcolor{teal}{A}$\mapsto$\strue, and one for \textcolor{teal}{A}$\mapsto$\sfalse.
These are, respectively, \strue $||$ \textcolor{red}{B} and \sfalse $||$ \textcolor{red}{B}.
After constant folding and strength reduction, we have \strue and \textcolor{red}{B}.

\begin{figure*}
    \includegraphics[width=1.00\linewidth]{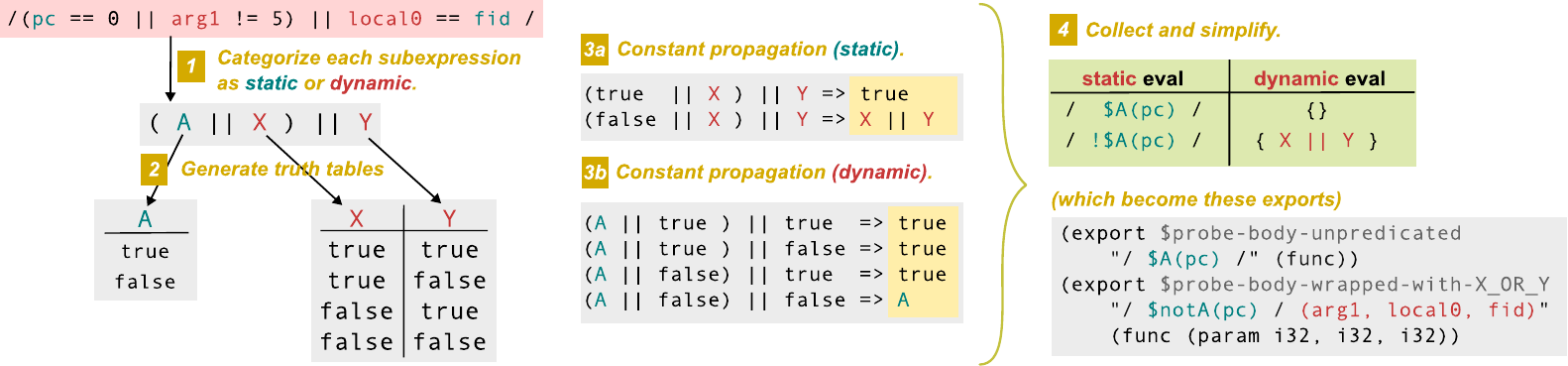}
    \caption{Illustration of splitting a mixed predicate (upper left) into subexpressions to evaluate statically and dynamically (right-side output), reducing predication overhead in the engine.}
    \label{fig:pred-opt-engine}
\end{figure*}

While symbolic approaches might apply, an effective brute-force strategy is:
\begin{enumerate}
    \item Categorize each subexpression as static or dynamic.
    \item Generate a truth table for the outcome of each static subexpression.
    \item Substitute constants, fold, and strength-reduce to produce residual dynamic predicates.
    \item Collect results, combine equivalences, and simplify.
\end{enumerate}
Figure~\ref{fig:pred-opt-engine} shows an in-depth example of this static analysis.
After optimization, the \ourmonitor contains two specialized probes for the single DSL probe expression.
One probe is purely statically predicated, while the second probe handles the case where the static predicate is \sfalse, requiring the dynamic residual to be evaluated.

\subsection{Engine Optimizations}\label{subsubsec:engine-intrinsics}

The Wizard engine supports dynamic instrumentation by allowing user-defined probe objects to be attached as specific Wasm instructions.
These probes are implemented in the engine (i.e. as host code) and as we've seen, form the basis of our \ourprobe implementation.
This section details three key optimizations to reduce overheads associated with this mechanism. 

To understand the runtime overhead of the engine-probe mechanism, consider the definition of a Wizard probe.
Such probes are defined in terms of Virgil classes with a \texttt{fire()} method that is called at runtime.
The method receives a (function, \texttt{pc}) pair and a handle to a lazily-allocated reflective \texttt{FrameAccessor} object for accessing the operand stack, locals, etc.
Flexible instrumentation is seamlessly supported in both the interpreter and JIT tiers of the engine but requires a relatively complicated transfer sequence from application code to probe code, consisting of these steps:
\begin{enumerate}
    \item commit program state to the machine frame
    \item save interpreter or JIT VM state, such as a value stack pointer
    \item look up the probe object through indirections from the instance pointer
    \item invoke the (host) probe function
\end{enumerate}

Depending on the complexity of the probe's logic, it may incur additional overheads as it executes:

\begin{enumerate}
  \item potentially redundant computation in the probe function
  \item materializing the \texttt{FrameAccessor} object if requested
  \item expensive frame access via indirection through the \texttt{FrameAccessor}
\end{enumerate}

In practice, most probes don't require the full generality of the probing system, allowing for engine optimizations that remove some of the above steps at probe callsites.
Removing unneeded work in the transfer sequence and making accessing of program variables cheaper are key to reducing instrumentation overhead.
This is especially true for the \ourprobe, which implements an adapter that transfers data from application Wasm context, through the runtime, then back to another Wasm context.
This is relatively expensive to do via the engine APIs.
In the general case, each \ourlang function may have both constants and variables as \ourargs; the probe loads their values from the \texttt{FrameAccessor} and calls the target function with the \ourargs using the engine's host-to-Wasm call mechanism.
A host-to-Wasm call has relatively high overhead, as it may allocate a fresh machine stack\footnote{as part of Wizard's support for the \texttt{stack-switching} proposal} for the call and the \ourargs for the call are copied to and from the heap at least once.
To reduce overhead of \ourprobe instrumentation, we identified three optimizations:

\textbf{1. (Interpreter) Use trampolines for customized transfer and a Wasm-to-Wasm call.}
Since the \ourprobe object is an engine-level wrapper around a Wasm function, the obvious optimization is to avoid the expensive transfer sequence detailed above and simply make the application-to-instrumentation transfer a regular Wasm call.
The only wrinkle is that \ourargs must be passed somehow.
Since these are known at binding time and consist of a mix of constants and frame accesses, we wrote a \emph{trampoline} generator which produces a small machine code stub for each \ourprobe signature, at match time, before program execution.
The trampoline contains the minimum machine instructions specialized to each unique signature that performs the necessary moves to/from the operand stack slots to properly load \ourargs.
It then performs a Wasm-to-Wasm call to the \ourprobe function, landing either in JIT-compiled code for the function or back in the interpreter, with a new frame for the probe function.
After the call returns, the trampoline dispatches on the original application opcode to continue execution.
One trampoline is generated for each unique \ourprobe \emph{signature}, so many probe objects sharing the same signature also share trampolines, incurring minimum space and time overhead.
This optimization completely removes the need to materialize a \texttt{FrameAccessor} or setup an entirely new execution context with the \emph{Execute API}.



\textbf{2. (JIT) Replace the call to the \ourprobe with a direct Wasm call in the compiler.}
Wizard employs a single-pass compiler~\cite{WizardJit} (SPC) as its JIT tier.
This compiler already contains support for intrinsifying counter and top-of-stack probes~\cite{WizardInstr}.
We extend the compiler with an additional case for when a probe at a given site is attached to a \ourprobe function.
Like interpretation, we can benefit from specializing the transfer sequence, but further, the compiler can \emph{inline} this transfer sequence.
As the SPC allocates registers and tracks constants via abstract interpretation, the inlined transfer sequence is more efficient.
Moreover, the JIT can emit a direct call to the \ourprobe function.

\textbf{3. (JIT) Inline \emph{compiled} \ourmonitor functions directly into application code.}
We can further reduce overhead for probes by inlining the probe's Wasm code into application code at probe sites.
This avoids the cost of a Wasm function call---a disproportionate overhead when considering many probe functions are short sequences that should be no more than a couple machine instructions.
Inlining short probe functions is a big win that brings fully programmable instrumentation up to speed with handwritten intrinsification.
For comparison, Wizard's prior \texttt{CountProbe} intrinsification allowed using simple counter probes at the cost of just two inline machine instructions per probe site.
Expressing a counter as a \ourprobe with six Wasm instructions to load, increment, and store an internal memory address results in the JIT compiler inlining just five machine instructions\footnote{One instruction to load the memory base of the probe module, one to put a match-time constant into a register, one to load, one to add, and one to store. With some folding and better instruction selection in an optimizing compiler, this could be a single instruction.}.

We chose to implement very simple\footnote{In our implementation, only straight-line probe functions are inlined.} inlining in the SPC as a proof of concept.
As a state-of-the-art optimizing baseline compiler, the SPC performs abstract interpretation and models the control stack during a single forward compilation pass.
When reaching a \ourprobe, we modify the SPC to simply push a new control stack entry modeling the entry of a new function, juggle a bit of abstract state, and then continue to generate code from the bytecode of the inlined function.
Since the SPC already models constants and allocates registers, the new bytecode is inlined \emph{and optimized} in the context of the application code.
In particular, locations of \ourargs are statically known at the site and any results from match-time callbacks become \emph{constants}.
This means that match-time callbacks that are typically used to allocate a metadata entry will have their result inlined as a constant into the probe code at each different probe site.
This approach makes good use of the SPC context's register allocation state and abstract stack state to remove redundant memory operations and moves, but requires some additional slots in the machine frame.
There are some minor details such as the metadata for stack walking to correctly reconstruct two frames from a single frame, but this is no different or worse than handling inlining of application functions.
The total addition to the compiler amounts to 250 lines of code, most of which deals with variable transfers.
We do admit that inlining in an optimizing compiler would be considerably more elegant, and maintainers of engines with two compiler tiers might consider only implementing inlining (of instrumentation) in the optimizing tier.

\section{Evaluation}\label{sec:evaluation}

In this section, we answer the following research questions:
\begin{enumerate}
    \item \textbf{Section~\ref{subsec:programmability}}: Does \ourlang provide a general-purpose framework for programmable instrumentation?
    \item \textbf{Section~\ref{subsec:programmability}}: Does the syntax of \ourlang result in more usable and maintainable monitor implementations than the state-of-the-art?
    \item \textbf{Sections~\ref{subsubsec:int-opts} and~\ref{subsubsec:jit-opts}}: Do the JIT and interpreter optimizations reasonably boost performance?
    \item \textbf{Sections~\ref{subsubsec:vs-fmks} and~\ref{subsubsec:vs-pin}}: Is the instrumentation performance competitive with state-of-the art Wasm instrumentation and traditionally performant frameworks?
\end{enumerate}

\begin{table}[ht]
    \tiny
    \centering
    \begin{tabular}{p{5cm} | p{2cm} | p{5cm}}
        Monitor & Framework & Lines of Code (LoC) \\
        \hline\hline
        \multirow{4}{5cm}{\hangindent=1em \texttt{branches}: profiles the branches taken during program execution for the \texttt{select}, \texttt{br}, \texttt{br\_if}, and \texttt{br\_table} opcodes.
        \footnote{The garbage collection proposal's branching opcodes are not currently supported in this version of the monitor, but will be added to later iterations.}} & \cellcolor{yellow!25} \ourlang & \cellcolor{yellow!25} 25 \\ \cline{2-3}
        & Wasabi & 29 \\ \cline{2-3}
        & Walrus / \orca & 291 \\ \cline{2-3}
        & Virgil & 221 \\ \cline{2-3}
        \hline
        \multirow{4}{5cm}{\hangindent=1em \texttt{imix}: assigns categories (e.g.\ \emph{control}, \emph{load}, etc) to each opcode and counts the dynamic occurrences of each category.} & \cellcolor{yellow!25} \ourlang (our work) & \cellcolor{yellow!25} 299 \\ \cline{2-3}
        & Wasabi & 84 \\ \cline{2-3}
        & Walrus / \orca & 441 \\ \cline{2-3}
        & Virgil & 705 \\ \cline{2-3}
        \hline
        \multirow{4}{5cm}{\hangindent=1em \texttt{hotness}: counts the number of times each individual instruction is executed.} & \cellcolor{yellow!25} \ourlang (our work) & \cellcolor{yellow!25} 4 \\ \cline{2-3}
        & Wasabi & 96 \\ \cline{2-3}
        & Walrus / \orca & 242 \\ \cline{2-3}
        & Virgil & 189 \\ \cline{2-3}
        \hline
        \multirow{4}{5cm}{\hangindent=1em \texttt{cache\_sim}: simulates a cache lookup on each memory load and store opcode and reports the total cache hits and misses.} & \cellcolor{yellow!25} \ourlang (our work) & \cellcolor{yellow!25} 13 (plus 228 for cache library code) \\ \cline{2-3}
        & Wasabi & 168 \\ \cline{2-3}
        & Walrus / \orca & 234 \\ \cline{2-3}
        & Virgil & 295 \\ \cline{2-3}
        \hline
    \end{tabular}
    \caption{The suite of monitors implemented for this evaluation.}
    \label{tab:loc}
\end{table}

\subsection{Programmability and Usability}\label{subsec:programmability}
For our evaluation, we implement a diverse set of monitors (Table~\ref{tab:loc}) to demonstrate that \ourlang provides concise programmable instrumentation with high performance.
Through specializing \ourlang syntax to commonly-needed operations and storage primitives, fewer lines of code are needed.
For example, it only requires 4 lines of code for a simple \texttt{hotness} monitor and 13 lines of code for the relatively complicated \texttt{cache\_sim} monitor.
In contrast, manual bytecode rewriting with Walrus/\orca require tedious handwritten bytecode and careful management of instrumentation metadata.
For more general programmability, \ourlang scripts can rely on libraries.
This greatly simplifies complex monitoring problems.
For example, a succinct implementation of the \texttt{imix} monitor can be written in 4 lines of code, shown below.

\begin{lstlisting}[escapechar=!]
!\hspace{20pt}\textcolor{keyword}{report var}! dyn_categories: !\textcolor{special}{map}<\textcolor{special}{u32}, \textcolor{special}{i32}!>;
!\hspace{20pt}\textcolor{keyword}{wasm}:\textcolor{keyword}{opcode}:\textcolor{special}{*}:\textcolor{keyword}{before}! {
    !\hspace{20pt}!dyn_categories[category_id]++; !\textcolor{comment}{// category_id is a bound variable}!
!\hspace{20pt}!}
\end{lstlisting}

The above is short, but it has poor performance due to the use of a library-provided \texttt{map} data structure.
\ourlang is flexible enough to support multiple implementations of a monitor, optimized to the desired criteria (conciseness, performance, maintainability) of the user.
A faster version of the \texttt{imix} monitor could use one directive per opcode:

\begin{lstlisting}[escapechar=!]
!\hspace{20pt}\textcolor{keyword}{report var} control: \textcolor{special}{u32};!
!\hspace{20pt}\textcolor{keyword}{report var} arith: \textcolor{special}{u32};!
!\hspace{20pt}!...
!\hspace{20pt}\textcolor{keyword}{wasm}:\textcolor{keyword}{opcode}:\textcolor{keyword}{br\_if}:\textcolor{keyword}{before} \{ control++; \}!
!\hspace{20pt}\textcolor{keyword}{wasm}:\textcolor{keyword}{opcode}:\textcolor{keyword}{i32.add}:\textcolor{keyword}{before} \{ arith++; \}!
!\hspace{20pt}!...
\end{lstlisting}

\subsection{Performance}\label{subsec:performance}

To evaluate the performance of monitoring with \ourlang, we evaluate overhead costs on the Polybench/C benchmark suite~\cite{PolyBench} with the \texttt{medium} dataset.
We evaluate our system against the state of the art Pin~\cite{Pin} machine-code-rewriting instrumentation framework, by compiling the Wasm binaries back to C using \texttt{wasm2c}, then to  \texttt{x86-64} using \texttt{gcc}.
To obtain baseline (uninstrumented) execution times (see Table~\ref{tab:baselines} in the appendix), we ran each benchmark on Wizard interpreter/JIT tiers, natively, and on V8 using the \texttt{{-}{-}liftoff-only} option, and averaged execution times from five runs.
We used (patched) Wizard version 25$\beta$.2717, V8 version 13.6.0, \texttt{wasm2c} version 1.0.34, Pin version 3.31, and \texttt{gcc} version 13.2.0 with optimization level \texttt{-O2}.
All experiments were performed on a machine with two Xeon Platinum 8168 CPUs $@$ 2.70GHz with 384GiB of DDR4 RAM.
Given uninstrumented execution time in seconds, $T_{u}$, and instrumented execution time, $T_{i}$, we use the term \emph{relative overhead} to refer to the ratio $T_{i}/T_{u}$, while \emph{absolute overhead} refers to $T_{i}-T_{u}$.

\subsubsection{\ourlang vs.\ State-of-the-Art Wasm Frameworks}\label{subsubsec:vs-fmks}

\begin{figure*}
    \includegraphics[width=1.00\linewidth]{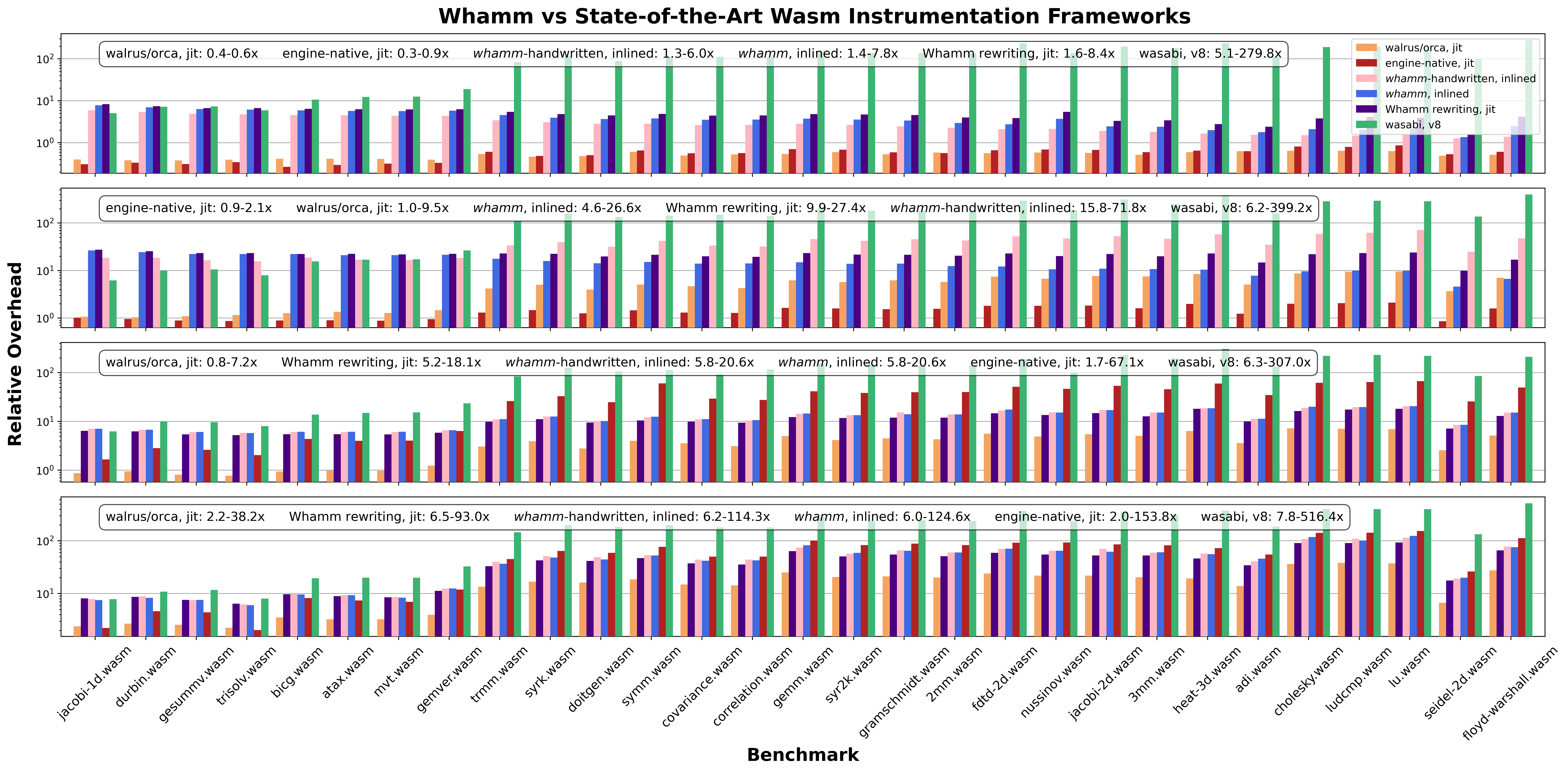}
    \caption{The \emph{relative execution} times, log scale, of the \texttt{branches}, \texttt{hotness}, \texttt{imix}, and \texttt{cache\_sim} monitors against state-of-the-art Wasm instrumentation frameworks across the Polybench/C benchmark suite sorted by absolute execution time.
    Ratios are relative to uninstrumented execution time.
    \emph{Lower is better.}
    }
    \label{fig:whamm-vs-fmk-all-monitors}
\end{figure*}

We compare the \emph{relative overhead} imposed by \ourlang rewriting and \ourmonitor{s} with state-of-the-art WebAssembly instrumentation frameworks by implementing reasonably-equivalent monitors in Native Wizard instrumentation, Wasabi~\cite{Wasabi}, IR injection with the \orca and Walrus libraries~\cite{Walrus}, and handwritten \ourmonitor{s}, see Figure~\ref{fig:whamm-vs-fmk-all-monitors}.
We ran each of these implementations in their most-optimized configuration on the Wizard JIT.
As Wasabi requires an engine that runs JavaScript, we ran the Wasabi experiments on the V8 engine and limited execution to its baseline compiler, Liftoff, for a fair comparison\footnote{Experiment results demonstrate that Wasabi instrumentation overhead is dominated by the context switching between JS and Wasm execution, not the compiler tier.}.
The uninstrumented time for each benchmark can be found in Table~\ref{tab:baselines}.

\textbf{\ourlang vs. Wasabi.}
Wasabi rewrites Wasm bytecode to inject runtime calls to JavaScript instrumentation.
While this is known to be expensive for Wasabi, our results show that \oureng instrumentation is similarly expensive when implemented as runtime calls.
Both impose orders of magnitude overhead, shown in results for Wasabi and our unoptimized \texttt{int-rt-*}/\texttt{jit-rt-*} configurations in Figures~\ref{fig:int-opts} and~\ref{fig:jit-opts-all-monitors}.

\textbf{\ourlang vs. Walrus/\orca.}
Directly injecting into Wasm IR with library support allows one to inject custom, nearly optimal, bytecode via rewriting.
For short-running programs, our data shows that the binding and setup time dominate, giving the advantage to Walrus/\orca and native.
However, for long-running programs, all monitors show both \ourlang targets to be competitive with manual rewriting, usually within $1.5-2\times$.
Another difference is that the Walrus/\orca monitors use more optimized handwritten bytecode injection that has denser data representations, and do not actually report results.
Improvements to the \ourcomp could lessen this gap.

\textbf{\ourlang vs. Native Wizard Instrumentation.}
Native Wizard instrumentation is highly optimized for special cases via JIT intrinsics that recognize special \texttt{Probe} definitions in the engine API.
For example, the \texttt{CountProbe} optimizes the emitted machine code to just two instructions, and the top-of-stack probe directly passes the top-of-stack to probe callbacks.
The impact of such optimizations is seen in the efficiency of the \texttt{hotness} and \texttt{branches} monitors.
Native engine instrumentation is faster than \ourlang on small benchmarks for the \texttt{cache\_sim} monitor, partly due to the overhead of JIT inlining for \ourmonitor{s} and stack manipulations for the \ourlang rewriting target.
Note that the native top-of-stack specialization duplicates the semantics of a \ourmonitor requesting the \texttt{arg0} variable.
While our engine interface does not yet beat the performance of these highly specialized probes, it is still competitive, and as we discuss in future work, more compiler optimization can narrow this gap.

\textbf{\ourlang vs. Handwritten Monitor Modules.}
We demonstrate the effectiveness of the \ourcomp by implementing logically-equivalent handwritten {\ourmonitor}s designed for maximum performance.
While the handwritten \texttt{branches} monitor is more performant than that of \ourlang for short-running programs (mostly load time effects), for long-running programs \ourlang monitors have nearly-equivalent performance as the handwritten bytecode.

\subsubsection{\oureng Interpreter Optimization}\label{subsubsec:int-opts}
\begin{figure*}
    \includegraphics[width=1.00\linewidth]{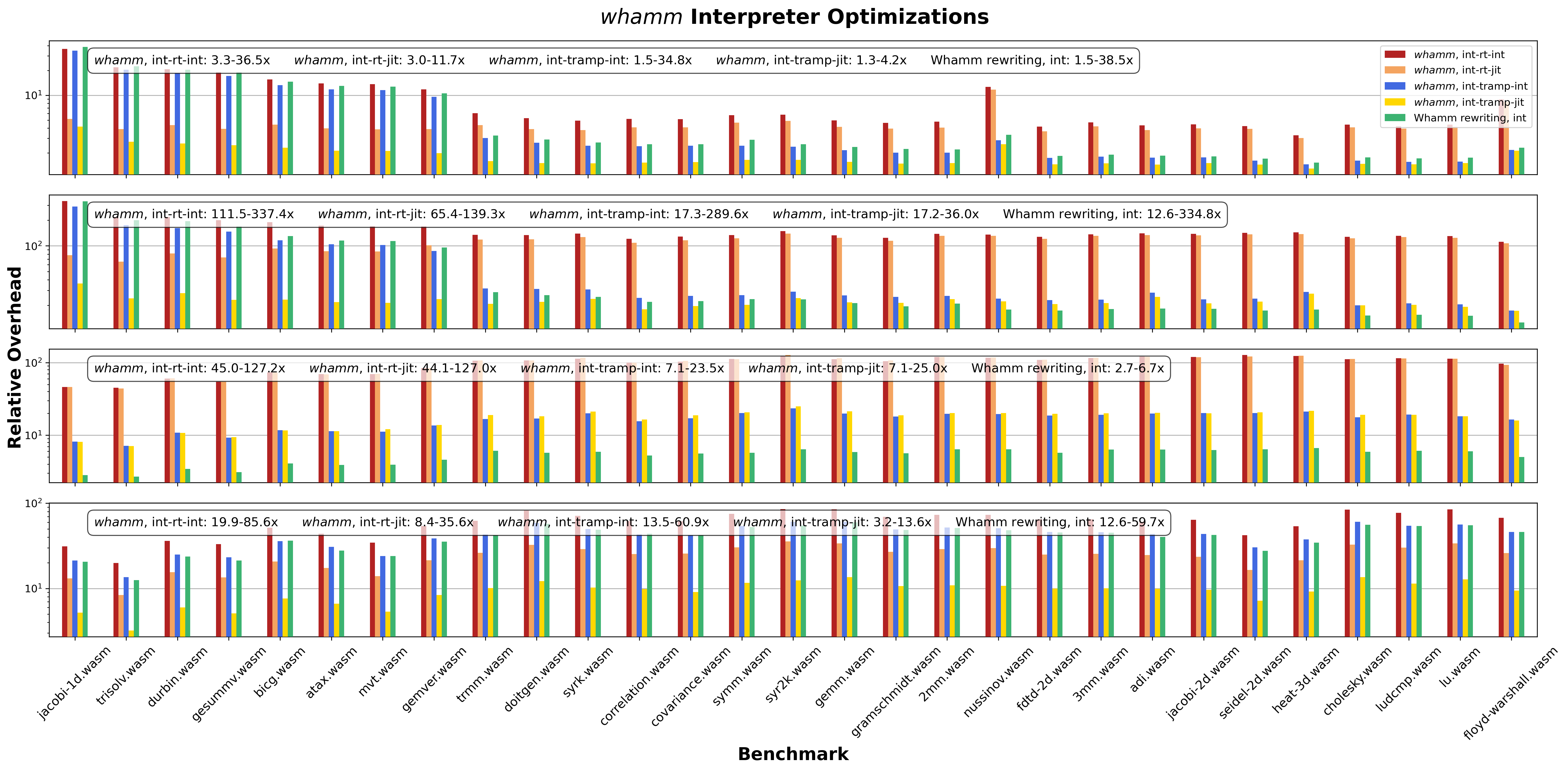}
    \caption{
        The \emph{relative execution} times, log scale, of the \texttt{branches}, \texttt{hotness}, \texttt{imix}, and \texttt{cache\_sim} monitors when instrumenting an interpreted program.
        We compare the base unoptimized case, \emph{int-rt-int}, JITing the instrumentation of the unoptimized case, \emph{int-rt-jit}, the interpreter trampoline optimization, \emph{int-tramp-int}, and interpreting simple \ourlang rewriting.
        Ratios are relative to uninstrumented execution time.
        \emph{Lower is better.}}
    \label{fig:int-opts}
\end{figure*}

We evaluate the effect of the \emph{trampoline} interpreter optimization in Figure~\ref{fig:int-opts}.
In the base case, \emph{int-rt-int}, the application code is interpreted; when an instrumented point is reached, a runtime call is made to invoke the \ourfunc.
The instrumentation code itself is then interpreted.
We show the impact of simply JITing the instrumentation code, leaving the runtime call as-is in the configuration \emph{int-rt-jit}.
The optimized case, \emph{int-tramp-int}, leverages the \emph{trampoline} that enables a direct call to the \ourfunc, skipping the runtime call.
Further optimization is then shown with the \emph{int-tramp-jit} configuration, in which the \emph{trampoline} calls a JITed \ourfunc.
We further show the performance of simply interpreting an application rewritten with \ourlang.

For the \texttt{branches} and \texttt{hotness} monitors, trampoline generation imposes more overhead than runtime calls into JITed probes for short-running programs.
However, as base program runtime increases, the runtime call dominates the overhead (consistent with the Wasabi framework).
For the \texttt{imix} monitor, the trampoline saves the expensive runtime call, but lags behind rewriting performance.
This is because the trampoline still has an indirection overhead, whereas rewriting is inlined in the application code.
As the \texttt{cache\_sim} probe bodies are more complex than other monitors, JIT-compiling instrumentation provides a large performance boost as the overhead is dominated by actual instrumentation logic rather than the injection and invocation strategies.

\subsubsection{\oureng JIT Optimization.}\label{subsubsec:jit-opts}
\begin{figure*}
    \includegraphics[width=1.00\linewidth]{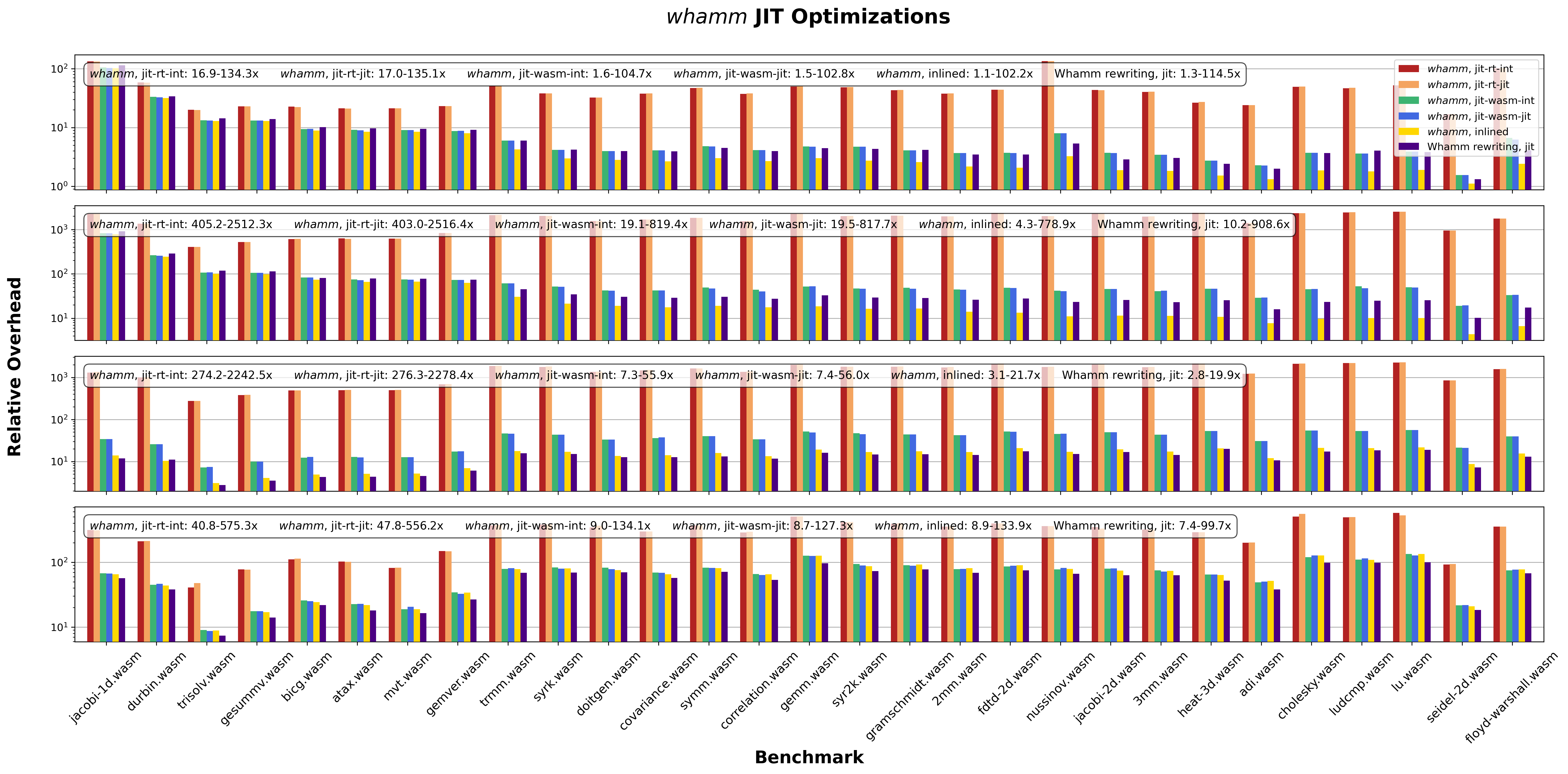}
    \caption{
        The \emph{relative execution} times, log scale, of the \texttt{branches}, \texttt{hotness}, \texttt{imix}, and \texttt{cache\_sim} monitors when instrumenting a JITed program.
        We compare the base unoptimized case, \emph{jit-rt-int}, JITing the instrumentation of the unoptimized case, \emph{jit-rt-jit}, enabling the JIT wasm-to-wasm call optimization, \emph{jit-wasm-int}, JITing the instrumentation of that case, \emph{jit-wasm-jit}, \emph{inlining} the instrumentation by the JIT, and JITing simple \ourlang rewriting.
        Ratios are relative to uninstrumented execution time.
        \emph{Lower is better.}
    }
    \label{fig:jit-opts-all-monitors}
\end{figure*}

We evaluate the \oureng JIT optimizations in Figure~\ref{fig:jit-opts-all-monitors}.
In the base case, \emph{jit-rt-int}, the application code is JIT-compiled; when an instrumented point is reached, a runtime call is made to invoke the probe callback, the instrumentation code is then interpreted.
We show the impact of simply JIT-compiling the instrumentation code, leaving the runtime call as-is in the configuration, \emph{jit-rt-jit}.
In the first optimized case, \emph{jit-wasm-int}, the application code performs a direct Wasm call at the probe site in the compiled machine code, skipping the runtime call.
The \emph{inlined} case further optimizes probe invocation by directly inlining the \ourfunc at the probe site.
We further show the performance of simply JIT-compiling an application rewritten with \ourlang.

In general, instrumentation overhead is dominated by the runtime call, if present.
Removing this call overhead results in a large performance boost, as shown by \texttt{jit-wasm-int}.
Further, inlining reduces the overhead costs to be even lower than rewriting!
The \texttt{branches} monitor requires the use of \texttt{arg0} to check the dynamic target of a branching opcode.
As Wasm is a stack machine, accessing the top-of-stack when rewriting requires stack manipulations (see Figure~\ref{fig:injection-rewriting}), while the engine knows the value's register.
This observation optimizes \emph{transfer} time for inlined probe bodies.
Inlining the \texttt{imix} monitors has similar performance to \ourlang rewriting, regardless of base program run time.
For the \texttt{cache\_sim} monitor, we observe that, after removing the runtime call, the overhead is dominated by the time spent in the cache library, making optimization techniques for transfer cost negligible.

\subsubsection{\ourlang vs.\ Pin, a Traditional Highly Performant Framework}\label{subsubsec:vs-pin}

We evaluate the performance of the \texttt{imix}, \texttt{icount}, and \texttt{cache\_sim} monitors in \ourlang against logically-equivalent implementations in Pin~\cite{Pin}, a highly-performant machine-code rewriting instrumentation framework.
We compare the \emph{absolute overhead}, $T_{i}-T_{u}$, imposed by instrumenting \emph{machine code} with Pin and \emph{JITed Wasm bytecode} with \ourlang rewriting and \oureng.
As shown in Table~\ref{tab:baselines}, the base performance of machine code is, on average, $5\times$ faster than Wizard's baseline JIT compiler.
As we are comparing \emph{absolute overhead}, the difference in base runtime is factored out, but instrumentation code compiled in Pin has an advantage.

\begin{figure*}
    \includegraphics[width=.80\linewidth]{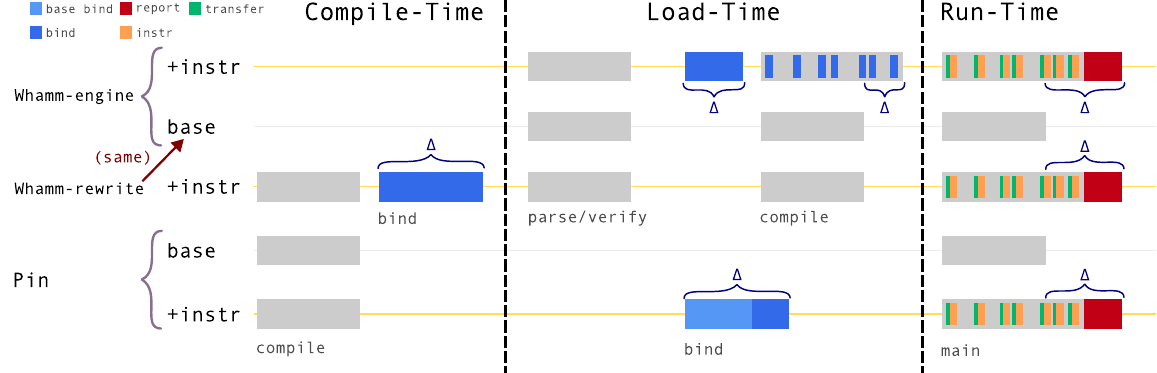}
    \caption{
        Visualization of the various overheads imposed by Pin, \ourlang rewriting, and \ourmonitor{s}.
        The \texttt{base} rows represent the runtime for uninstrumented machine code or bytecode.
        Rewriting imposes some static overhead to inject instrumentation into the program, \emph{bind} cost.
        This cost is paid at runtime for both Pin and \ourmonitor{s}.
        Pin requires some \emph{base bind} overhead to start up the framework itself plus some cost relative to the size of the program.
        All three frameworks have overhead for \emph{transfer}-ing into monitor code, executing it, (\emph{instr}) and flushing monitor state (\emph{report}).
    }
    \label{fig:overheads}
\end{figure*}

\begin{figure*}
    \includegraphics[width=1.00\linewidth]{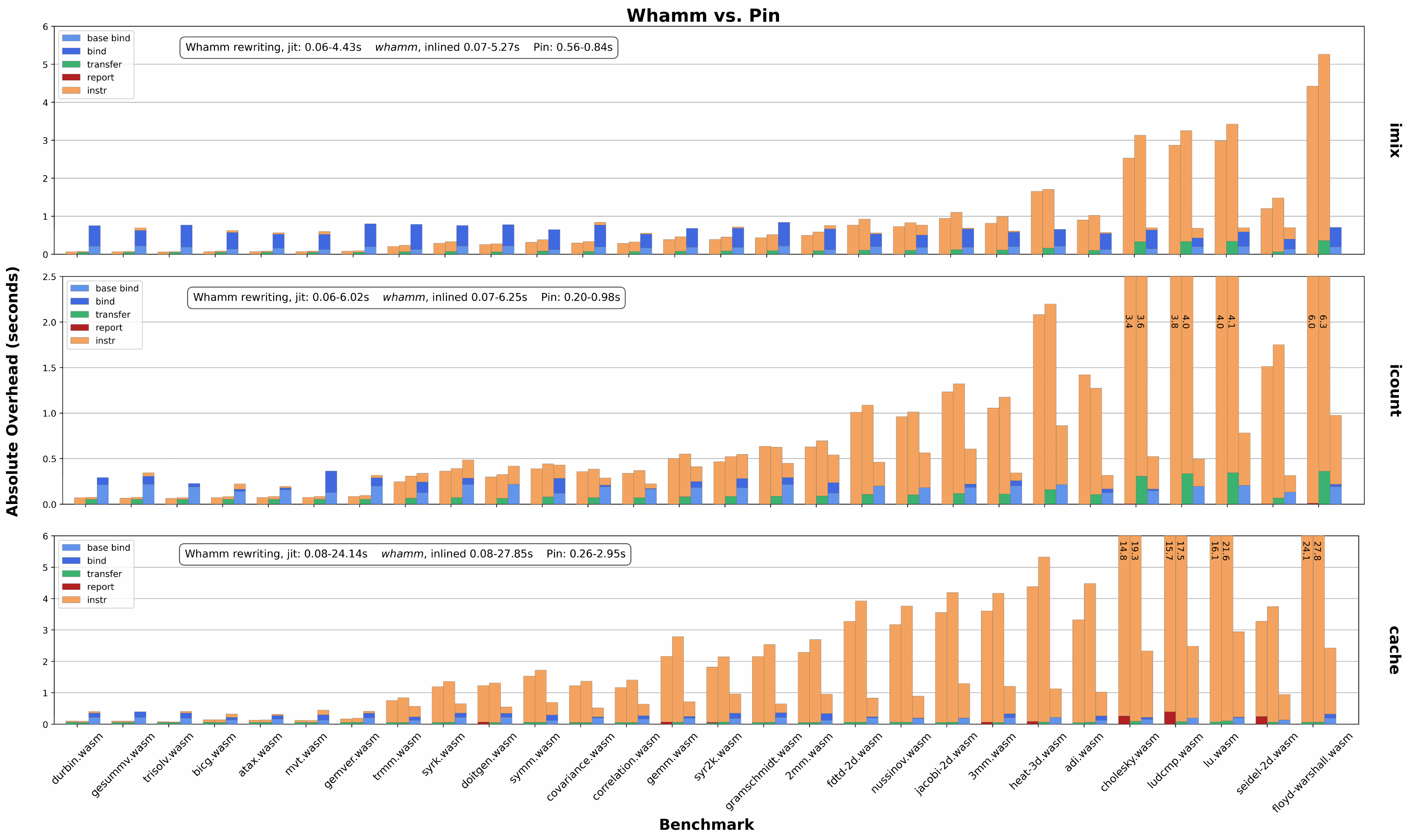}
    \caption{
        The \emph{absolute overhead} of the \texttt{imix}, \texttt{icount}, and \texttt{cache\_sim} monitors when using the \ourlang rewriting (left column), the \ourlang engine interface (middle column), and Pin framework (right column) broken down by type of overhead (see Figure~\ref{fig:overheads}).
        \ourlang Rewriting performance is computed for a JITed instrumented application and \ourlang engine interface is run is its most-optimized configuration, JIT inlining of probes.
        Numbers are relative to uninstrumented execution time.
        \emph{Lower is better.}
    }
    \label{fig:whamm-vs-pin-all}
\end{figure*}

Figure~\ref{fig:overheads} explains the various types of overheads imposed by each framework and compares the time required for each in Figure~\ref{fig:whamm-vs-pin-all}.
We calculate the \emph{bind} overhead with engine metrics for \oureng (done statically in rewriting), the \emph{transfer} overhead by instrumenting a program with empty probes, the \emph{report} overhead through engine metrics for \oureng and measuring the difference between instrumentation overhead with and without flushing for \ourlang rewriting, and \emph{instr} as the rest of the time.
We calculate \emph{base bind} for Pin by starting up the framework for each benchmark with no monitor activated and \emph{bind} through taking the difference between the overhead of instrumenting with empty probes and \emph{base bind}.
We were unable to compute the \emph{transfer} and \emph{report} overheads for Pin, so these are not separated from the \emph{instr} overhead in the charts.

For shorter running programs, especially for simple probes like those in \texttt{imix} and \texttt{icount}, the \ourlang rewriting and engine strategies are more performant than Pin due to their small, or non-existent, bind overheads.
The \texttt{cache\_sim} monitor overhead becomes large for all long-running benchmarks because these programs, except \texttt{seidel-2d}, all have a high number of memory read/write operations resulting in heavy instrumentation.
While Pin is fastest for long-running programs, it clearly has large, fixed startup overhead and surprisingly large binding time.
A significant portion of the gap for long-running programs is the code quality of Wizard's SPC JIT when compiling complex monitor code.
We expect that in another engine with a more optimizing compiler, monitor code speed will approach optimal.
Of note is that our performance was achievable with very few lines of code for all monitors (see Table~\ref{tab:loc}) that is both intuitive and maintainable rather than having to write the low-level C code necessary for Pin instrumentation.


\section{Related work}\label{sec:related-work}

Dynamic analysis frameworks have been the subject of a vast amount of research.
The main choices include whether to inject code into the source code or application binary, recompile programs with new instrumentation, or emulate the process space of a program.
These approaches have various tradeoffs including performance overheads, intrusiveness concerns, usability, and composibility.

\textbf{Code injection}
A common approach is to inject monitoring code directly into a program binary ~\cite{BinRewriteSurvey}, either by injecting the entire monitoring code \emph{inline} or with a \emph{trampoline} that jumps to out-of-line instrumentation code.

Code injection can be performed statically, that is, separately from execution.
For Java bytecode, early static tools include Soot~\cite{Soot} and Bloat~\cite{NystromThesis}.
As Aspect-Oriented Programming (AOP) emerged, so did tools DiSL~\cite{DISL}, AspectJ~\cite{AspectJ} and BISM~\cite{BISM, EfficientInstrumentation} that targeted \emph{joinpoints}.
While AOP instrumentation was readable, it was unable to manipulate bytecode at a low-level.
The ASM~\cite{ASM} framework is capable of low-level manipulation, but was considered user unfriendly.
The later Clojure~\cite{Clojure} bytecode manipulation library enables low-level manipulations in a readable, expressive, and concise notation.

Alternatively, code can be injected at runtime. DTrace~\cite{DTrace,DTrace2}, inspired by Paradyn~\cite{Paradyn}, enables tracing at both the user and kernel layer of the OS with dynamically-injected trampolines.
Dyninst~\cite{Dyninst} interfaces with a program's CFG and maps modifications to concrete binary rewrites.
A user can tie instrumentation to instructions or CFG abstractions (e.g. function entry/exit).
Recent research in this direction~\cite{ToggleProbe,Odin,TowardMinimalMonitoring} focuses on reducing instrumentation overhead with low-level optimizations.

For Wasm, many solutions exist like~\cite{SEISMIC}, Walrus~\cite{Walrus}, BREWasm~\cite{brewasm}, and Wasabi~\cite{Wasabi}, but none are as user-friendly as the bytecode rewriting framework presented in this work.

\textbf{Recompilation.}
Compiled programs can be \emph{recompiled} to inject code using several techniques.
For example, compiled code can be \emph{lifted} to a higher-level IR, instrumented, then recompiled.
Early examples of static lifting for instrumentation include ATOM~\cite{AtomTools} and EEL~\cite{EEL}.
Etch~\cite{Etch}, through observing an initial program execution, discovered dynamic program properties to inform static instrumentation.
DynamoRIO~\cite{DynamoRIO} and Pin~\cite{Pin} use dynamic recompilation of native binaries for instrumentation.
Their JIT compilers are purpose-built for instrumentation but since it runs code as part of the initial process, it is susceptible to being intrusive.

\textbf{Emulation.}
QEMU~\cite{QEMU} is a widely-used CPU emulator that virtualizes a user-space process while supporting non-intrusive instrumentation.
Valgrind~\cite{Valgrind}, primarily used as a memory debugger, is similar.
They include JIT compilers that are not necessarily ``purpose-built'' for instrumentation, but for cross-compilation, leading to an inherent overhead over virtual machines.

\textbf{Direct engine support.}
Virtual machines and their runtime systems can be designed with specific support for instrumentation.
In .NET~\cite{DotNetProfiling}, users build profiler DLLs that are loaded by the CLR into the same process as a target application and interface through callbacks.
The JVM Tool Interface~\cite{JvmTI} instruments Java bytecode by attaching callbacks to \emph{agents} written against the lower-level engine API.
However, evidence suggests that event handling in agents can be prohibitively slow and the engine API cannot handle events at the level of individual bytecodes.
For Wasm, the Wizard Research Engine~\cite{WizardEngine} provides support for monitoring by writing engine-specific probes in the Virgil language~\cite{WizardInstr}.
Our work leverages the existing support in the Wizard Engine to deliver easy-to-implement, yet portable engine-level instrumentation.

\textbf{Composability of analyses.}
Composability of instrumentation allows programmers to use multiple analyses simultaneously on the same application.
Some instrumentation frameworks have attempted to promote interoperability of monitors through meta-modeling~\cite{TowardInteroperability} and interface standardization~\cite{OMIS}.
Other frameworks have emphasized the need for composable monitors in works such as the built-in Wizard VM instrumentation support~\cite{WizardInstr} and a technique called polymorphic instrumentation~\cite{PolymorphicInstrumentation}.
Virtual-machine based solutions like the JVM, .NET, and the Wizard engine allow general composability.
Our engine monitoring interface goes further and enforces separation and composability of monitors by sandboxing them in Wasm modules.

\section{Conclusion and Future Work}

In this work, we presented \ourlang, a declarative instrumentation language.
The syntax of this language provides a semantically rich environment for natural instrumentation expression.
We showed that \ourlang's performance is competitive with performant frameworks due to novel optimizations, freeing users from writing low-level, tedious code.

In future work, we plan to expand the instrumentable events in \ourlang to support use cases such as security policy enforcement, modeling data flow between Wasm components, and resilience engineering tooling.
We also hope to aid further in the debuggability of instrumentation through creating support frameworks such as an IDE plugin for live programming with \ourlang and a test suite framework.
Another area of interest is designing a Wasm reflection and instrumentation API to enable the monitors themselves to perform the static bytecode matching logic for dynamic instrumentation.
We also have plans to further optimize \ourlang performance through implementing predicate optimizations as outlined in Section~\ref{subsubsec:predicate-opts}.
In this work we focused effort on prototyping in Wizard, which has advanced instrumentation, but no optimizing compiler.
Since such a compiler would presumably be good at optimizing Wasm code, it would excel at optimizing application and monitor code together, particularly as probe bodies are usually short but may have many common subexpressions across match sites.
Some instrumentation tasks might involve predicates that change infrequently, such as guarding detailed instrumentation for a rare program event.
While the underlying engine supports dynamically removing and adding new probes one-by-one, each individual insertion/removal can deoptimize a function.
Rather than insert many probes at once and then remove them all at once, instrumentation might be better expressed with predicates on global variables with the engine employing optimization guards~\cite{Graal}.
Our intention with \oureng was to entice production engines into supporting a standardized monitoring API by limiting the additional mechanisms needed, e.g. by reusing Wasm modules deftly.

\section{Data-Availability Statement}

%
%
%
%

We plan to submit an artifact for evaluation.
This artifact will include the benchmark suites used, scripts to run the experiments, usage instructions, and our experiment results as CSVs.

\begin{acks}
This work is supported by the WebAssembly Research Center and the NSF Grant Award \#2148301.
We would like to thank Adam Bratschi Kaye at Dfinity, Yan Chen at Fastly, Jeff Charles and Saúl Cabrera at Shopify, and Ulan Degenbaev.
\end{acks}

\bibliographystyle{ACM-Reference-Format}
\bibliography{paper}


\begin{thebibliography}{44}


\ifx \showCODEN    \undefined \def \showCODEN     #1{\unskip}     \fi
\ifx \showDOI      \undefined \def \showDOI       #1{#1}\fi
\ifx \showISBNx    \undefined \def \showISBNx     #1{\unskip}     \fi
\ifx \showISBNxiii \undefined \def \showISBNxiii  #1{\unskip}     \fi
\ifx \showISSN     \undefined \def \showISSN      #1{\unskip}     \fi
\ifx \showLCCN     \undefined \def \showLCCN      #1{\unskip}     \fi
\ifx \shownote     \undefined \def \shownote      #1{#1}          \fi
\ifx \showarticletitle \undefined \def \showarticletitle #1{#1}   \fi
\ifx \showURL      \undefined \def \showURL       {\relax}        \fi
\providecommand\bibfield[2]{#2}
\providecommand\bibinfo[2]{#2}
\providecommand\natexlab[1]{#1}
\providecommand\showeprint[2][]{arXiv:#2}

\bibitem[Fas(2020)]%
        {FastlyEdge}
 \bibinfo{year}{2020}\natexlab{}.
\newblock \bibinfo{title}{The edge of the multi-cloud}.
\newblock
  \bibinfo{howpublished}{\url{https://www.fastly.com/cassets/6pk8mg3yh2ee/79dsHLTEfYIMgUwVVllaa4/5e5330572b8f317f72e16696256d8138/WhitePaper-Multi-Cloud.pdf}}.
\newblock
\urldef\tempurl%
\url{https://www.fastly.com/cassets/6pk8mg3yh2ee/79dsHLTEfYIMgUwVVllaa4/5e5330572b8f317f72e16696256d8138/WhitePaper-Multi-Cloud.pdf}
\showURL{%
\tempurl}
\newblock
\shownote{(Accessed 2021-07-06)}.


\bibitem[Jvm(2021)]%
        {JvmTI}
 \bibinfo{year}{2021}\natexlab{}.
\newblock \bibinfo{title}{Java {V}irtual {M}achine {T}ools {I}nterface}.
\newblock
  \bibinfo{howpublished}{\url{https://docs.oracle.com/javase/8/docs/technotes/guides/jvmti/}}.
\newblock
\urldef\tempurl%
\url{https://docs.oracle.com/javase/8/docs/technotes/guides/jvmti/}
\showURL{%
\tempurl}
\newblock
\shownote{(Accessed 2021-07-29)}.


\bibitem[Wal(2023)]%
        {Walrus}
 \bibinfo{year}{2023}\natexlab{}.
\newblock \bibinfo{title}{Walrus: A {W}eb{A}ssembly transformation library}.
\newblock \bibinfo{howpublished}{\url{https://github.com/rustwasm/walrus}}.
\newblock
\urldef\tempurl%
\url{https://github.com/rustwasm/walrus}
\showURL{%
\tempurl}


\bibitem[Authors(2021)]%
        {DotNetProfiling}
\bibfield{author}{\bibinfo{person}{.NET~Wiki Authors}.}
  \bibinfo{year}{2021}\natexlab{}.
\newblock \bibinfo{title}{The {.NET} {P}rofiling {API}}.
\newblock
  \bibinfo{howpublished}{\url{https://learn.microsoft.com/en-us/dotnet/framework/unmanaged-api/profiling/profiling-overview}}.
\newblock
\urldef\tempurl%
\url{https://learn.microsoft.com/en-us/dotnet/framework/unmanaged-api/profiling/profiling-overview}
\showURL{%
\tempurl}
\newblock
\shownote{(Accessed 2023-8-4)}.


\bibitem[Bellard(2020)]%
        {QEMU}
\bibfield{author}{\bibinfo{person}{Fabrice Bellard}.}
  \bibinfo{year}{2020}\natexlab{}.
\newblock \bibinfo{title}{{QEMU}: A generic and open source machine emulator
  and virtualizer}.
\newblock \bibinfo{howpublished}{\url{http://qemu.org}}.
\newblock
\urldef\tempurl%
\url{http://qemu.org}
\showURL{%
\tempurl}
\newblock
\shownote{(Accessed 2023-8-07)}.


\bibitem[Bernat and Miller(2011)]%
        {Dyninst}
\bibfield{author}{\bibinfo{person}{Andrew~R. Bernat} {and}
  \bibinfo{person}{Barton~P. Miller}.} \bibinfo{year}{2011}\natexlab{}.
\newblock \showarticletitle{Anywhere, Any-Time Binary Instrumentation}. In
  \bibinfo{booktitle}{\emph{Proceedings of the 10th ACM SIGPLAN-SIGSOFT
  Workshop on Program Analysis for Software Tools}} (Szeged, Hungary)
  \emph{(\bibinfo{series}{PASTE '11})}. \bibinfo{publisher}{Association for
  Computing Machinery}, \bibinfo{address}{New York, NY, USA},
  \bibinfo{pages}{9–16}.
\newblock
\showISBNx{9781450308496}
\urldef\tempurl%
\url{https://doi.org/10.1145/2024569.2024572}
\showDOI{\tempurl}


\bibitem[Bruening et~al\mbox{.}(2003)]%
        {DynamoRIO}
\bibfield{author}{\bibinfo{person}{D Bruening}, \bibinfo{person}{T Garnett},
  {and} \bibinfo{person}{S Amarasinghe}.} \bibinfo{year}{2003}\natexlab{}.
\newblock \showarticletitle{An infrastructure for adaptive dynamic
  optimization}. In \bibinfo{booktitle}{\emph{International Symposium on Code
  Generation and Optimization, 2003. {CGO} 2003}} (San Francisco, CA, USA).
  \bibinfo{publisher}{IEEE Comput. Soc}.
\newblock


\bibitem[Bruneliere et~al\mbox{.}(2010)]%
        {TowardInteroperability}
\bibfield{author}{\bibinfo{person}{Hugo Bruneliere}, \bibinfo{person}{Jordi
  Cabot}, \bibinfo{person}{Cauê Clasen}, \bibinfo{person}{Frédéric Jouault},
  {and} \bibinfo{person}{Jean Bézivin}.} \bibinfo{year}{2010}\natexlab{}.
\newblock \showarticletitle{Towards Model Driven Tool Interoperability:
  Bridging Eclipse and Microsoft Modeling Tools}.
\newblock \bibinfo{journal}{\emph{Proceedings of the 6th European Conference on
  Modelling Foundations and Applications (ECMFA 2010)}}.
\newblock
\showISBNx{978-3-642-13594-1}
\urldef\tempurl%
\url{https://doi.org/10.1007/978-3-642-13595-8_5}
\showDOI{\tempurl}


\bibitem[Bruneton et~al\mbox{.}(2002)]%
        {ASM}
\bibfield{author}{\bibinfo{person}{Eric Bruneton}, \bibinfo{person}{Romain
  Lenglet}, {and} \bibinfo{person}{Thierry Coupaye}.}
  \bibinfo{year}{2002}\natexlab{}.
\newblock \showarticletitle{ASM: a code manipulation tool to implement
  adaptable systems}.
\newblock \bibinfo{journal}{\emph{Adaptable and extensible component systems}}
  \bibinfo{volume}{30}, \bibinfo{number}{19} (\bibinfo{year}{2002}).
\newblock


\bibitem[Cao et~al\mbox{.}(2023)]%
        {brewasm}
\bibfield{author}{\bibinfo{person}{Shangtong Cao}, \bibinfo{person}{Ningyu He},
  \bibinfo{person}{Yao Guo}, {and} \bibinfo{person}{Haoyu Wang}.}
  \bibinfo{year}{2023}\natexlab{}.
\newblock \showarticletitle{BREWasm: A General Static Binary Rewriting
  Framework for WebAssembly}. In \bibinfo{booktitle}{\emph{Static Analysis:
  30th International Symposium, SAS 2023, Cascais, Portugal, October 22–24,
  2023, Proceedings}} (Lisbon, Portugal). \bibinfo{publisher}{Springer-Verlag},
  \bibinfo{address}{Berlin, Heidelberg}, \bibinfo{pages}{139–163}.
\newblock
\showISBNx{978-3-031-44244-5}
\urldef\tempurl%
\url{https://doi.org/10.1007/978-3-031-44245-2_8}
\showDOI{\tempurl}


\bibitem[Chamith et~al\mbox{.}(2016)]%
        {ToggleProbe}
\bibfield{author}{\bibinfo{person}{Buddhika Chamith}, \bibinfo{person}{Bo~Joel
  Svensson}, \bibinfo{person}{Luke Dalessandro}, {and} \bibinfo{person}{Ryan~R.
  Newton}.} \bibinfo{year}{2016}\natexlab{}.
\newblock \showarticletitle{Living on the Edge: Rapid-Toggling Probes with
  Cross-Modification on X86}. In \bibinfo{booktitle}{\emph{Proceedings of the
  37th ACM SIGPLAN Conference on Programming Language Design and
  Implementation}} (Santa Barbara, CA, USA) \emph{(\bibinfo{series}{PLDI
  '16})}. \bibinfo{publisher}{Association for Computing Machinery},
  \bibinfo{address}{New York, NY, USA}, \bibinfo{pages}{16–26}.
\newblock
\showISBNx{9781450342612}
\urldef\tempurl%
\url{https://doi.org/10.1145/2908080.2908084}
\showDOI{\tempurl}


\bibitem[Cooper(2012)]%
        {DTrace2}
\bibfield{author}{\bibinfo{person}{Greg Cooper}.}
  \bibinfo{year}{2012}\natexlab{}.
\newblock \showarticletitle{{D}{T}race: Dynamic Tracing in {O}racle {S}olaris,
  {M}ac {OS} {X}, and {F}ree {BSD} by {B}rendan {G}regg and {J}im {M}auro}.
\newblock \bibinfo{journal}{\emph{SIGSOFT Softw. Eng. Notes}}
  \bibinfo{volume}{37}, \bibinfo{number}{1} (\bibinfo{date}{jan}
  \bibinfo{year}{2012}), \bibinfo{pages}{34}.
\newblock
\showISSN{0163-5948}
\urldef\tempurl%
\url{https://doi.org/10.1145/2088883.2088902}
\showDOI{\tempurl}


\bibitem[Gregg and Mauro(2011)]%
        {DTrace}
\bibfield{author}{\bibinfo{person}{Brendan Gregg} {and} \bibinfo{person}{Jim
  Mauro}.} \bibinfo{year}{2011}\natexlab{}.
\newblock \bibinfo{booktitle}{\emph{{D}{T}race: Dynamic Tracing in {O}racle
  {S}olaris, {M}ac {OS} {X} and {F}ree{BSD}} (\bibinfo{edition}{1st} ed.)}.
\newblock \bibinfo{publisher}{Prentice Hall Press}, \bibinfo{address}{USA}.
\newblock
\showISBNx{0132091518}


\bibitem[Haas et~al\mbox{.}(2017)]%
        {WasmPldi}
\bibfield{author}{\bibinfo{person}{Andreas Haas}, \bibinfo{person}{Andreas
  Rossberg}, \bibinfo{person}{Derek~L. Schuff}, \bibinfo{person}{Ben~L.
  Titzer}, \bibinfo{person}{Michael Holman}, \bibinfo{person}{Dan Gohman},
  \bibinfo{person}{Luke Wagner}, \bibinfo{person}{Alon Zakai}, {and}
  \bibinfo{person}{JF Bastien}.} \bibinfo{year}{2017}\natexlab{}.
\newblock \showarticletitle{Bringing the Web up to Speed with {W}eb{A}ssembly}.
  In \bibinfo{booktitle}{\emph{Proceedings of the 38th ACM SIGPLAN Conference
  on Programming Language Design and Implementation}} (Barcelona, Spain)
  \emph{(\bibinfo{series}{PLDI 2017})}. \bibinfo{publisher}{Association for
  Computing Machinery}, \bibinfo{address}{New York, NY, USA},
  \bibinfo{pages}{185–200}.
\newblock
\showISBNx{9781450349888}
\urldef\tempurl%
\url{https://doi.org/10.1145/3062341.3062363}
\showDOI{\tempurl}


\bibitem[Kiczales et~al\mbox{.}(2001)]%
        {AspectJ}
\bibfield{author}{\bibinfo{person}{Gregor Kiczales}, \bibinfo{person}{Erik
  Hilsdale}, \bibinfo{person}{Jim Hugunin}, \bibinfo{person}{Mik Kersten},
  \bibinfo{person}{Jeffrey Palm}, {and} \bibinfo{person}{William~G. Griswold}.}
  \bibinfo{year}{2001}\natexlab{}.
\newblock \showarticletitle{An Overview of {A}spect{J}}. In
  \bibinfo{booktitle}{\emph{Proceedings of the 15th European Conference on
  Object-Oriented Programming}} \emph{(\bibinfo{series}{ECOOP '01})}.
  \bibinfo{publisher}{Springer-Verlag}, \bibinfo{address}{Berlin, Heidelberg},
  \bibinfo{pages}{327–353}.
\newblock
\showISBNx{3540422064}


\bibitem[Larus and Schnarr(1995)]%
        {EEL}
\bibfield{author}{\bibinfo{person}{James~R. Larus} {and} \bibinfo{person}{Eric
  Schnarr}.} \bibinfo{year}{1995}\natexlab{}.
\newblock \showarticletitle{{EEL}: Machine-Independent Executable Editing}. In
  \bibinfo{booktitle}{\emph{Proceedings of the ACM SIGPLAN 1995 Conference on
  Programming Language Design and Implementation}} (La Jolla, California, USA)
  \emph{(\bibinfo{series}{PLDI '95})}. \bibinfo{publisher}{Association for
  Computing Machinery}, \bibinfo{address}{New York, NY, USA},
  \bibinfo{pages}{291–300}.
\newblock
\showISBNx{0897916972}
\urldef\tempurl%
\url{https://doi.org/10.1145/207110.207163}
\showDOI{\tempurl}


\bibitem[Lehmann and Pradel(2019)]%
        {Wasabi}
\bibfield{author}{\bibinfo{person}{Daniel Lehmann} {and}
  \bibinfo{person}{Michael Pradel}.} \bibinfo{year}{2019}\natexlab{}.
\newblock \showarticletitle{Wasabi: A Framework for Dynamically Analyzing
  {W}eb{A}ssembly}. In \bibinfo{booktitle}{\emph{Proceedings of the
  Twenty-Fourth International Conference on Architectural Support for
  Programming Languages and Operating Systems}} (Providence, RI, USA)
  \emph{(\bibinfo{series}{ASPLOS '19})}. \bibinfo{publisher}{Association for
  Computing Machinery}, \bibinfo{address}{New York, NY, USA},
  \bibinfo{pages}{1045–1058}.
\newblock
\showISBNx{9781450362405}
\urldef\tempurl%
\url{https://doi.org/10.1145/3297858.3304068}
\showDOI{\tempurl}


\bibitem[Ludwig and Wism{\"u}ller(1997)]%
        {OMIS}
\bibfield{author}{\bibinfo{person}{Thomas Ludwig} {and} \bibinfo{person}{Roland
  Wism{\"u}ller}.} \bibinfo{year}{1997}\natexlab{}.
\newblock \showarticletitle{OMIS 2.0 --- a universal interface for monitoring
  systems}. In \bibinfo{booktitle}{\emph{Recent Advances in Parallel Virtual
  Machine and Message Passing Interface}},
  \bibfield{editor}{\bibinfo{person}{Marian Bubak}, \bibinfo{person}{Jack
  Dongarra}, {and} \bibinfo{person}{Jerzy Wa{\'{s}}niewski}} (Eds.).
  \bibinfo{publisher}{Springer Berlin Heidelberg}, \bibinfo{address}{Berlin,
  Heidelberg}, \bibinfo{pages}{267--276}.
\newblock
\showISBNx{978-3-540-69629-2}


\bibitem[Luk et~al\mbox{.}(2005)]%
        {Pin}
\bibfield{author}{\bibinfo{person}{Chi-Keung Luk}, \bibinfo{person}{Robert
  Cohn}, \bibinfo{person}{Robert Muth}, \bibinfo{person}{Harish Patil},
  \bibinfo{person}{Artur Klauser}, \bibinfo{person}{Geoff Lowney},
  \bibinfo{person}{Steven Wallace}, \bibinfo{person}{Vijay~Janapa Reddi}, {and}
  \bibinfo{person}{Kim Hazelwood}.} \bibinfo{year}{2005}\natexlab{}.
\newblock \showarticletitle{Pin: Building Customized Program Analysis Tools
  with Dynamic Instrumentation}.
\newblock \bibinfo{journal}{\emph{SIGPLAN Not.}} \bibinfo{volume}{40},
  \bibinfo{number}{6} (\bibinfo{date}{June} \bibinfo{year}{2005}),
  \bibinfo{pages}{190–200}.
\newblock
\showISSN{0362-1340}
\urldef\tempurl%
\url{https://doi.org/10.1145/1064978.1065034}
\showDOI{\tempurl}


\bibitem[Marek et~al\mbox{.}(2012)]%
        {DISL}
\bibfield{author}{\bibinfo{person}{Luk\'{a}\v{s} Marek}, \bibinfo{person}{Alex
  Villaz\'{o}n}, \bibinfo{person}{Yudi Zheng}, \bibinfo{person}{Danilo
  Ansaloni}, \bibinfo{person}{Walter Binder}, {and} \bibinfo{person}{Zhengwei
  Qi}.} \bibinfo{year}{2012}\natexlab{}.
\newblock \showarticletitle{{D}i{SL}: A Domain-Specific Language for Bytecode
  Instrumentation}. In \bibinfo{booktitle}{\emph{Proceedings of the 11th Annual
  International Conference on Aspect-Oriented Software Development}} (Potsdam,
  Germany) \emph{(\bibinfo{series}{AOSD '12})}. \bibinfo{publisher}{Association
  for Computing Machinery}, \bibinfo{address}{New York, NY, USA},
  \bibinfo{pages}{239–250}.
\newblock
\showISBNx{9781450310925}
\urldef\tempurl%
\url{https://doi.org/10.1145/2162049.2162077}
\showDOI{\tempurl}


\bibitem[Miller et~al\mbox{.}(1995)]%
        {Paradyn}
\bibfield{author}{\bibinfo{person}{Barton~P. Miller}, \bibinfo{person}{Mark~D.
  Callaghan}, \bibinfo{person}{Jonathan~M. Cargille},
  \bibinfo{person}{Jeffrey~K. Hollingsworth}, \bibinfo{person}{R.~Bruce Irvin},
  \bibinfo{person}{Karen~L. Karavanic}, \bibinfo{person}{Krishna
  Kunchithapadam}, {and} \bibinfo{person}{Tia Newhall}.}
  \bibinfo{year}{1995}\natexlab{}.
\newblock \showarticletitle{The {P}aradyn Parallel Performance Measurement
  Tool}.
\newblock \bibinfo{journal}{\emph{Computer}} \bibinfo{volume}{28},
  \bibinfo{number}{11} (\bibinfo{date}{nov} \bibinfo{year}{1995}),
  \bibinfo{pages}{37–46}.
\newblock
\showISSN{0018-9162}
\urldef\tempurl%
\url{https://doi.org/10.1109/2.471178}
\showDOI{\tempurl}


\bibitem[Moret et~al\mbox{.}(2011)]%
        {PolymorphicInstrumentation}
\bibfield{author}{\bibinfo{person}{Philippe Moret}, \bibinfo{person}{Walter
  Binder}, {and} \bibinfo{person}{\'{E}ric Tanter}.}
  \bibinfo{year}{2011}\natexlab{}.
\newblock \showarticletitle{Polymorphic bytecode instrumentation}. In
  \bibinfo{booktitle}{\emph{Proceedings of the Tenth International Conference
  on Aspect-Oriented Software Development}} (Porto de Galinhas, Brazil)
  \emph{(\bibinfo{series}{AOSD '11})}. \bibinfo{publisher}{Association for
  Computing Machinery}, \bibinfo{address}{New York, NY, USA},
  \bibinfo{pages}{129–140}.
\newblock
\showISBNx{9781450306058}
\urldef\tempurl%
\url{https://doi.org/10.1145/1960275.1960292}
\showDOI{\tempurl}


\bibitem[Munsters et~al\mbox{.}(2021)]%
        {Oron}
\bibfield{author}{\bibinfo{person}{A\"{a}ron Munsters}, \bibinfo{person}{Angel
  Luis~Scull Pupo}, \bibinfo{person}{Jim Bauwens}, {and}
  \bibinfo{person}{Elisa~Gonzalez Boix}.} \bibinfo{year}{2021}\natexlab{}.
\newblock \showarticletitle{Oron: Towards a Dynamic Analysis Instrumentation
  Platform for {A}ssembly{S}cript}. In \bibinfo{booktitle}{\emph{Companion
  Proceedings of the 5th International Conference on the Art, Science, and
  Engineering of Programming}} (Cambridge, United Kingdom)
  \emph{(\bibinfo{series}{Programming '21})}. \bibinfo{publisher}{Association
  for Computing Machinery}, \bibinfo{address}{New York, NY, USA},
  \bibinfo{pages}{6–13}.
\newblock
\showISBNx{9781450389860}
\urldef\tempurl%
\url{https://doi.org/10.1145/3464432.3464780}
\showDOI{\tempurl}


\bibitem[Nakakaze et~al\mbox{.}(2022)]%
        {WasmIndMach}
\bibfield{author}{\bibinfo{person}{Otoya Nakakaze}, \bibinfo{person}{Istv{\'a}n
  Koren}, \bibinfo{person}{Florian Brillowski}, {and} \bibinfo{person}{Ralf
  Klamma}.} \bibinfo{year}{2022}\natexlab{}.
\newblock \showarticletitle{Retrofitting Industrial Machines with
  {W}eb{A}ssembly on the Edge}. In \bibinfo{booktitle}{\emph{Web Information
  Systems Engineering -- WISE 2022}},
  \bibfield{editor}{\bibinfo{person}{Richard Chbeir}, \bibinfo{person}{Helen
  Huang}, \bibinfo{person}{Fabrizio Silvestri}, \bibinfo{person}{Yannis
  Manolopoulos}, {and} \bibinfo{person}{Yanchun Zhang}} (Eds.).
  \bibinfo{publisher}{Springer International Publishing},
  \bibinfo{address}{Cham}, \bibinfo{pages}{241--256}.
\newblock
\showISBNx{978-3-031-20891-1}


\bibitem[Nethercote and Seward(2007)]%
        {Valgrind}
\bibfield{author}{\bibinfo{person}{Nicholas Nethercote} {and}
  \bibinfo{person}{Julian Seward}.} \bibinfo{year}{2007}\natexlab{}.
\newblock \showarticletitle{Valgrind: A Framework for Heavyweight Dynamic
  Binary Instrumentation}.
\newblock \bibinfo{journal}{\emph{SIGPLAN Not.}} \bibinfo{volume}{42},
  \bibinfo{number}{6} (\bibinfo{date}{jun} \bibinfo{year}{2007}),
  \bibinfo{pages}{89–100}.
\newblock
\showISSN{0362-1340}
\urldef\tempurl%
\url{https://doi.org/10.1145/1273442.1250746}
\showDOI{\tempurl}


\bibitem[Nystrom(1998)]%
        {NystromThesis}
\bibfield{author}{\bibinfo{person}{Nathaniel~John Nystrom}.}
  \bibinfo{year}{1998}\natexlab{}.
\newblock \emph{\bibinfo{title}{Bytecode-Level Analysis and Optimization of
  {Java} Classes}}.
\newblock \bibinfo{thesistype}{Master's\ thesis}. \bibinfo{school}{Purdue
  University}.
\newblock


\bibitem[Pouchet(2016)]%
        {PolyBench}
\bibfield{author}{\bibinfo{person}{Louis-No\"el Pouchet}.}
  \bibinfo{year}{2016}\natexlab{}.
\newblock \bibinfo{title}{{P}oly{B}ench}.
\newblock
\newblock
\urldef\tempurl%
\url{https://sourceforge.net/projects/polybench/}
\showURL{%
\tempurl}


\bibitem[Reichelt et~al\mbox{.}(2023)]%
        {TowardMinimalMonitoring}
\bibfield{author}{\bibinfo{person}{David~Georg Reichelt},
  \bibinfo{person}{Stefan K\"{u}hne}, {and} \bibinfo{person}{Wilhelm
  Hasselbring}.} \bibinfo{year}{2023}\natexlab{}.
\newblock \showarticletitle{Towards Solving the Challenge of Minimal Overhead
  Monitoring}. In \bibinfo{booktitle}{\emph{Companion of the 2023 ACM/SPEC
  International Conference on Performance Engineering}} (Coimbra, Portugal)
  \emph{(\bibinfo{series}{ICPE '23 Companion})}.
  \bibinfo{publisher}{Association for Computing Machinery},
  \bibinfo{address}{New York, NY, USA}, \bibinfo{pages}{381–388}.
\newblock
\showISBNx{9798400700729}
\urldef\tempurl%
\url{https://doi.org/10.1145/3578245.3584851}
\showDOI{\tempurl}


\bibitem[Rodrigues and Barreiros(2022)]%
        {AspectWasm}
\bibfield{author}{\bibinfo{person}{João Rodrigues} {and}
  \bibinfo{person}{Jorge Barreiros}.} \bibinfo{year}{2022}\natexlab{}.
\newblock \showarticletitle{Aspect-Oriented {W}eb{A}ssembly Transformation}. In
  \bibinfo{booktitle}{\emph{2022 17th Iberian Conference on Information Systems
  and Technologies (CISTI)}}. \bibinfo{pages}{1--6}.
\newblock
\urldef\tempurl%
\url{https://doi.org/10.23919/CISTI54924.2022.9820136}
\showDOI{\tempurl}


\bibitem[Romer et~al\mbox{.}(1997)]%
        {Etch}
\bibfield{author}{\bibinfo{person}{Ted Romer}, \bibinfo{person}{Geoff Voelker},
  \bibinfo{person}{Dennis Lee}, \bibinfo{person}{Alec Wolman},
  \bibinfo{person}{Wayne Wong}, \bibinfo{person}{Hank Levy},
  \bibinfo{person}{Brian Bershad}, {and} \bibinfo{person}{Brad Chen}.}
  \bibinfo{year}{1997}\natexlab{}.
\newblock \showarticletitle{Instrumentation and Optimization of {W}in32/{I}ntel
  Executables Using {E}tch}. In \bibinfo{booktitle}{\emph{Proceedings of the
  USENIX Windows NT Workshop on The USENIX Windows NT Workshop 1997}} (Seattle,
  Washington) \emph{(\bibinfo{series}{NT'97})}. \bibinfo{publisher}{USENIX
  Association}, \bibinfo{address}{USA}, \bibinfo{pages}{1}.
\newblock


\bibitem[Sen et~al\mbox{.}(2013)]%
        {Jalangi}
\bibfield{author}{\bibinfo{person}{Koushik Sen}, \bibinfo{person}{Swaroop
  Kalasapur}, \bibinfo{person}{Tasneem Brutch}, {and} \bibinfo{person}{Simon
  Gibbs}.} \bibinfo{year}{2013}\natexlab{}.
\newblock \showarticletitle{Jalangi: A Selective Record-Replay and Dynamic
  Analysis Framework for {J}ava{S}cript}. In
  \bibinfo{booktitle}{\emph{Proceedings of the 2013 9th Joint Meeting on
  Foundations of Software Engineering}} (Saint Petersburg, Russia)
  \emph{(\bibinfo{series}{ESEC/FSE 2013})}. \bibinfo{publisher}{Association for
  Computing Machinery}, \bibinfo{address}{New York, NY, USA},
  \bibinfo{pages}{488–498}.
\newblock
\showISBNx{9781450322379}
\urldef\tempurl%
\url{https://doi.org/10.1145/2491411.2491447}
\showDOI{\tempurl}


\bibitem[Soueidi et~al\mbox{.}(2020)]%
        {BISM}
\bibfield{author}{\bibinfo{person}{Chukri Soueidi}, \bibinfo{person}{Ali
  Kassem}, {and} \bibinfo{person}{Yli{\`e}s Falcone}.}
  \bibinfo{year}{2020}\natexlab{}.
\newblock \showarticletitle{{BISM}: bytecode-level instrumentation for software
  monitoring}. In \bibinfo{booktitle}{\emph{Runtime Verification: 20th
  International Conference, RV 2020, Los Angeles, CA, USA, October 6--9, 2020,
  Proceedings 20}}. Springer, \bibinfo{pages}{323--335}.
\newblock


\bibitem[Soueidi et~al\mbox{.}(2023)]%
        {EfficientInstrumentation}
\bibfield{author}{\bibinfo{person}{Chukri Soueidi}, \bibinfo{person}{Marius
  Monnier}, {and} \bibinfo{person}{Yli\`{e}s Falcone}.}
  \bibinfo{year}{2023}\natexlab{}.
\newblock \showarticletitle{Efficient and expressive bytecode-level
  instrumentation for Java programs}.
\newblock \bibinfo{journal}{\emph{Int. J. Softw. Tools Technol. Transf.}}
  \bibinfo{volume}{25}, \bibinfo{number}{4} (\bibinfo{date}{jun}
  \bibinfo{year}{2023}), \bibinfo{pages}{453–479}.
\newblock
\showISSN{1433-2779}
\urldef\tempurl%
\url{https://doi.org/10.1007/s10009-023-00708-z}
\showDOI{\tempurl}


\bibitem[Srivastava and Eustace(1994)]%
        {AtomTools}
\bibfield{author}{\bibinfo{person}{Amitabh Srivastava} {and}
  \bibinfo{person}{Alan Eustace}.} \bibinfo{year}{1994}\natexlab{}.
\newblock \showarticletitle{{ATOM}: A System for Building Customized Program
  Analysis Tools}. \bibinfo{publisher}{Association for Computing Machinery},
  \bibinfo{address}{New York, NY, USA}.
\newblock
\showISBNx{089791662X}
\urldef\tempurl%
\url{https://doi.org/10.1145/178243.178260}
\showDOI{\tempurl}


\bibitem[Titzer(2021)]%
        {WizardEngine}
\bibfield{author}{\bibinfo{person}{Ben~L. Titzer}.}
  \bibinfo{year}{2021}\natexlab{}.
\newblock \bibinfo{title}{{W}izard, {A}n advanced {W}ebAssembly {E}ngine for
  {R}esearch}.
\newblock
  \bibinfo{howpublished}{\url{https://github.com/titzer/wizard-engine}}.
\newblock
\urldef\tempurl%
\url{https://github.com/titzer/wizard-engine}
\showURL{%
\tempurl}
\newblock
\shownote{(Accessed 2021-07-29)}.


\bibitem[Titzer(2024)]%
        {WizardJit}
\bibfield{author}{\bibinfo{person}{Ben~L. Titzer}.}
  \bibinfo{year}{2024}\natexlab{}.
\newblock \showarticletitle{Whose Baseline Compiler is it Anyway?}
  \emph{(\bibinfo{series}{CGO '24})}. \bibinfo{publisher}{Association for
  Computing Machinery}, \bibinfo{address}{New York, NY, USA}.
\newblock


\bibitem[Titzer et~al\mbox{.}(2024)]%
        {WizardInstr}
\bibfield{author}{\bibinfo{person}{Ben~L. Titzer}, \bibinfo{person}{Elizabeth
  Gilbert}, \bibinfo{person}{Bradley Wei~Jie Teo}, \bibinfo{person}{Yash
  Anand}, \bibinfo{person}{Kazuyuki Takayama}, {and} \bibinfo{person}{Heather
  Miller}.} \bibinfo{year}{2024}\natexlab{}.
\newblock \showarticletitle{Flexible Non-intrusive Dynamic Instrumentation for
  WebAssembly}. In \bibinfo{booktitle}{\emph{Proceedings of the 29th ACM
  International Conference on Architectural Support for Programming Languages
  and Operating Systems, Volume 3}} (La Jolla, CA, USA)
  \emph{(\bibinfo{series}{ASPLOS '24})}. \bibinfo{publisher}{Association for
  Computing Machinery}, \bibinfo{address}{New York, NY, USA},
  \bibinfo{pages}{398–415}.
\newblock
\showISBNx{9798400703867}
\urldef\tempurl%
\url{https://doi.org/10.1145/3620666.3651338}
\showDOI{\tempurl}


\bibitem[Umatani et~al\mbox{.}(2015)]%
        {Clojure}
\bibfield{author}{\bibinfo{person}{Seiji Umatani}, \bibinfo{person}{Tomoharu
  Ugawa}, {and} \bibinfo{person}{Masahiro Yasugi}.}
  \bibinfo{year}{2015}\natexlab{}.
\newblock \showarticletitle{Design and Implementation of a Java Bytecode
  Manipulation Library for Clojure}.
\newblock \bibinfo{journal}{\emph{Journal of Information Processing}}
  \bibinfo{volume}{23}, \bibinfo{number}{5} (\bibinfo{year}{2015}),
  \bibinfo{pages}{716--729}.
\newblock
\urldef\tempurl%
\url{https://doi.org/10.2197/ipsjjip.23.716}
\showDOI{\tempurl}


\bibitem[Vall\'{e}e-Rai et~al\mbox{.}(2010)]%
        {Soot}
\bibfield{author}{\bibinfo{person}{Raja Vall\'{e}e-Rai}, \bibinfo{person}{Phong
  Co}, \bibinfo{person}{Etienne Gagnon}, \bibinfo{person}{Laurie Hendren},
  \bibinfo{person}{Patrick Lam}, {and} \bibinfo{person}{Vijay Sundaresan}.}
  \bibinfo{year}{2010}\natexlab{}.
\newblock \showarticletitle{Soot: A {J}ava Bytecode Optimization Framework}. In
  \bibinfo{booktitle}{\emph{CASCON First Decade High Impact Papers}} (Toronto,
  Ontario, Canada) \emph{(\bibinfo{series}{CASCON '10})}.
  \bibinfo{publisher}{IBM Corp.}, \bibinfo{address}{USA},
  \bibinfo{pages}{214–224}.
\newblock
\urldef\tempurl%
\url{https://doi.org/10.1145/1925805.1925818}
\showDOI{\tempurl}


\bibitem[Varda({[n.\,d.]})]%
        {CloudflareWasm}
\bibfield{author}{\bibinfo{person}{Kenton Varda}.}
  \bibinfo{year}{[n.\,d.]}\natexlab{}.
\newblock \bibinfo{title}{{WebAssembly} on {C}loudflare {W}orkers}.
\newblock
  \bibinfo{howpublished}{\url{https://blog.cloudflare.com/webassembly-on-cloudflare-workers/}}.
\newblock
\urldef\tempurl%
\url{https://blog.cloudflare.com/webassembly-on-cloudflare-workers/}
\showURL{%
\tempurl}
\newblock
\shownote{(Accessed 2021-07-06)}.


\bibitem[Wang et~al\mbox{.}(2022)]%
        {Odin}
\bibfield{author}{\bibinfo{person}{Mingzhe Wang}, \bibinfo{person}{Jie Liang},
  \bibinfo{person}{Chijin Zhou}, \bibinfo{person}{Zhiyong Wu},
  \bibinfo{person}{Xinyi Xu}, {and} \bibinfo{person}{Yu Jiang}.}
  \bibinfo{year}{2022}\natexlab{}.
\newblock \showarticletitle{{O}din: On-Demand Instrumentation with on-the-Fly
  Recompilation}. In \bibinfo{booktitle}{\emph{Proceedings of the 43rd ACM
  SIGPLAN International Conference on Programming Language Design and
  Implementation}} (San Diego, CA, USA) \emph{(\bibinfo{series}{PLDI 2022})}.
  \bibinfo{publisher}{Association for Computing Machinery},
  \bibinfo{address}{New York, NY, USA}, \bibinfo{pages}{1010–1024}.
\newblock
\showISBNx{9781450392655}
\urldef\tempurl%
\url{https://doi.org/10.1145/3519939.3523428}
\showDOI{\tempurl}


\bibitem[Wang et~al\mbox{.}(2018)]%
        {SEISMIC}
\bibfield{author}{\bibinfo{person}{Wenhao Wang}, \bibinfo{person}{Benjamin
  Ferrell}, \bibinfo{person}{Xiaoyang Xu}, \bibinfo{person}{Kevin~W. Hamlen},
  {and} \bibinfo{person}{Shuang Hao}.} \bibinfo{year}{2018}\natexlab{}.
\newblock \showarticletitle{SEISMIC: SEcure In-lined Script Monitors for
  Interrupting Cryptojacks}. In \bibinfo{booktitle}{\emph{Computer Security:
  23rd European Symposium on Research in Computer Security, ESORICS 2018,
  Barcelona, Spain, September 3-7, 2018, Proceedings, Part II}} (Barcelona,
  Spain). \bibinfo{publisher}{Springer-Verlag}, \bibinfo{address}{Berlin,
  Heidelberg}, \bibinfo{pages}{122–142}.
\newblock
\showISBNx{978-3-319-98988-4}
\urldef\tempurl%
\url{https://doi.org/10.1007/978-3-319-98989-1_7}
\showDOI{\tempurl}


\bibitem[Wenzl et~al\mbox{.}(2019)]%
        {BinRewriteSurvey}
\bibfield{author}{\bibinfo{person}{Matthias Wenzl}, \bibinfo{person}{Georg
  Merzdovnik}, \bibinfo{person}{Johanna Ullrich}, {and} \bibinfo{person}{Edgar
  Weippl}.} \bibinfo{year}{2019}\natexlab{}.
\newblock \showarticletitle{From Hack to Elaborate Technique -- A Survey on
  Binary Rewriting}.
\newblock \bibinfo{journal}{\emph{ACM Comput. Surv.}} \bibinfo{volume}{52},
  \bibinfo{number}{3}, Article \bibinfo{articleno}{49} (\bibinfo{date}{jun}
  \bibinfo{year}{2019}), \bibinfo{numpages}{37}~pages.
\newblock
\showISSN{0360-0300}
\urldef\tempurl%
\url{https://doi.org/10.1145/3316415}
\showDOI{\tempurl}


\bibitem[W\"{u}rthinger et~al\mbox{.}(2013)]%
        {Graal}
\bibfield{author}{\bibinfo{person}{Thomas W\"{u}rthinger},
  \bibinfo{person}{Christian Wimmer}, \bibinfo{person}{Andreas W\"{o}\ss{}},
  \bibinfo{person}{Lukas Stadler}, \bibinfo{person}{Gilles Duboscq},
  \bibinfo{person}{Christian Humer}, \bibinfo{person}{Gregor Richards},
  \bibinfo{person}{Doug Simon}, {and} \bibinfo{person}{Mario Wolczko}.}
  \bibinfo{year}{2013}\natexlab{}.
\newblock \showarticletitle{One {VM} to {R}ule {T}hem {A}ll}. In
  \bibinfo{booktitle}{\emph{Proceedings of the 2013 ACM International Symposium
  on New Ideas, New Paradigms, and Reflections on Programming \& Software}}
  (Indianapolis, Indiana, USA) \emph{(\bibinfo{series}{Onward\! 2013})}.
  \bibinfo{publisher}{Association for Computing Machinery},
  \bibinfo{address}{New York, NY, USA}, \bibinfo{pages}{187–204}.
\newblock
\showISBNx{9781450324724}
\urldef\tempurl%
\url{https://doi.org/10.1145/2509578.2509581}
\showDOI{\tempurl}


\end{thebibliography}

\appendix
\begin{table}[ht]
    \centering
    \begin{tabular}{p{0.12\linewidth} | p{0.15\linewidth} | p{0.1\linewidth} | p{0.1\linewidth} | p{0.1\linewidth} | p{0.17\linewidth}}
        Suite & Benchmark & Wizard INT (s) & Wizard JIT (s) & V8 (s) & Machine Code, \texttt{wasm2c} (s) \\
        \hline\hline
        polybench & 2mm.wasm & 0.515 & 0.046 & 0.027 & 0.013 \\
        polybench & 3mm.wasm & 0.903 & 0.070 & 0.033 & 0.023\\
        polybench & adi.wasm & 0.942 & 0.102 & 0.061 & 0.041 \\
        polybench & atax.wasm & 0.029 & 0.013 & 0.010 & 0.001 \\
        polybench & bicg.wasm & 0.027 & 0.013 & 0.011 & 0.002 \\
        polybench & cholesky.wasm & 3.017 & 0.169 & 0.090 & 0.067 \\
        polybench & correlation.wasm & 0.305 & 0.033 & 0.017 & 0.008 \\
        polybench & covariance.wasm & 0.307 & 0.034 & 0.017 & 0.008 \\
        polybench & doitgen.wasm & 0.226 & 0.030 & 0.015 & 0.006 \\
        polybench & durbin.wasm & 0.017 & 0.010 & 0.001 & 0.001 \\
        polybench & fdtd-2d.wasm & 0.876 & 0.057 & 0.025 & 0.018 \\
        polybench & floyd-warshall.wasm & 6.334 & 0.389 & 0.114 & 0.072 \\
        polybench & gemm.wasm & 0.414 & 0.033 & 0.019 & 0.009 \\
        polybench & gemver.wasm & 0.037 & 0.015 & 0.010 & 0.002 \\
        polybench & gesummv.wasm & 0.017 & 0.011 & 0.010 & 0.002 \\
        polybench & gramschmidt.wasm & 0.491 & 0.040 & 0.023 & 0.015 \\
        polybench & heat-3d.wasm & 1.570 & 0.096 & 0.043 & 0.030 \\
        polybench & jacobi-1d.wasm & 0.011 & 0.011 & 0.008 & 0.001 \\
        polybench & jacobi-2d.wasm & 1.019 & 0.068 & 0.031 & 0.018 \\
        polybench & ludcmp.wasm & 3.297 & 0.178 & 0.097 & 0.073 \\
        polybench & lu.wasm & 3.438 & 0.179 & 0.104 & 0.097 \\
        polybench & mvt.wasm & 0.029 & 0.015 & 0.010 & 0.001 \\
        polybench & nussinov.wasm & 0.746 & 0.060 & 0.036 & 0.012 \\
        polybench & seidel-2d.wasm & 1.351 & 0.202 & 0.102 & 0.104 \\
        polybench & symm.wasm & 0.324 & 0.034 & 0.020 & 0.009 \\
        polybench & syr2k.wasm & 0.380 & 0.036 & 0.020 & 0.011 \\
        polybench & syrk.wasm & 0.274 & 0.025 & 0.016 & 0.006 \\
        polybench & trisolv.wasm & 0.016 & 0.012 & 0.011 & 0.001 \\
        polybench & trmm.wasm & 0.189 & 0.021 & 0.014 & 0.005 \\
        \hline
    \end{tabular}
    \caption{The baseline execution time for the benchmark suites, uninstrumented.}
    \label{tab:baselines}
\end{table}

\setlength{\grammarindent}{8em} 
\begin{figure}
\begin{center}
\begin{grammar}
<probe-def> ::= <provider>:<package>:<event>:<mode> <predicate>? <block>

<predicate> ::= `/' <expr> `/'

<block> ::= `{' ( if-stmt | statement ) * `}'

<if-stmt> ::= `if' `(' <expr> `)' <block> ( <else-stmt> | <elif> )?

<else-stmt> ::= `else' <block>

<elif> ::= `elif' `(' <expr> `)' <block> ( <else-stmt> | <elif> )?

<statement> ::= <decl> `;'
	\alt <decl> `=' ( <ternary> | <expr> ) `;'
	\alt <call> `;'
	\alt ( <get-map> | "ID" ) `=' ( <ternary> | <expr> ) `;'
	\alt ( <get-map> | "ID" ) `++' `;'
	\alt ( <get-map> | "ID" ) `--' `;'
	\alt `return' <expr>? `;'

<decl> ::= <var-decorators> `var' "ID" `:' <type>

<var-decorators> ::= ( `report' | `unshared' )*

<call> ::= "ID" `.' "ID" `(' <arg>? ( `,' <arg> )* `)'
	\alt "ID" `(' <arg>? ( `,' <arg> )* `)'

<arg> ::= <expr> | <ternary>

<ternary> ::= <expr> `?' <expr> `:' <expr>

<expr> ::= <operand> <binop> <operand>
      \alt <unop> <operand>
      \alt <operand> `as' <type-primitive>
      \alt <call>
      \alt `(' <expr> `)'
      \alt <value>

<binop> ::= `<<' | `>>'
	| `&&' | `||'
	| `&' | `|' | `^'
	| `==' | `!=' | `>=' | `>' | `<=' | `<'
	| `+' | `-'
	| `*' | `/' | `

<unop> ::= `!' | `~'

<value> ::= "BOOL" | <get-map> | "ID" | "FLOAT" | "INT" | "STRING" | <tuple>

<get-map> ::= "ID" `[' <expr> `]'

<tuple> ::= `(' `)'
       \alt `(' <value> ( `,' <value> )* `)'

<type> ::= <type-primitive>
	| <type-tuple>
      	| <type-map>

<type-primitive> ::= `u8' | `i8'
      | `u16'
      | `i16'
      | `u32'
      | `i32'
      | `f32'
      | `u64'
      | `i64'
      | `f64'
      | `bool'
      | `str'

<type-tuple> ::= `(' `)'
	    \alt `(' <type> ( `,' <type> )* `)'

<type-map> ::= `map<' <type> `,` <type> `>'

\end{grammar}
\end{center}
\caption{Syntax of \ourlang.}
\end{figure}

\end{document}